\documentclass[11pt]{article}
\usepackage{jcappub}
\usepackage{bm}
\usepackage{color}
\usepackage{graphicx}
\newcommand{\be}{\begin{equation}}
\newcommand{\ee}{\end{equation}}
\newcommand{\een}{\end{subequations}}
\newcommand{\ben}{\begin{subequations}}
\newcommand{\beq}{\begin{eqalignno}}
\newcommand{\eeq}{\end{eqalignno}}

\newcommand{\lsim}{\mathrel{\mathop{\kern 0pt \rlap
      {\raise.2ex\hbox{$<$}}}\lower.9ex\hbox{\kern-.190em $ \sim$}}}
\newcommand{\gsim}{\mathrel{\mathop{\kern 0pt
      \rlap{\raise.2ex\hbox{$>$}}}\lower.9ex\hbox{\kern-.190em $\sim$}}}

\title{A systematic halo--independent analysis of direct detection
  data within the framework of Inelastic Dark Matter}

\author[a]{Stefano Scopel,}
\author[b]{Kook-Hyun Yoon}
\emailAdd{scopel@sogang.ac.kr}
\emailAdd{koreasds@naver.com}
\affiliation{Department of Physics, Sogang University, Seoul 121-742, South Korea}

\abstract{We present a systematic halo--independent analysis of
  available Weakly Interacting Massive Particles (WIMP) direct
  detection data within the framework of Inelastic Dark Matter (IDM).
  We show that, when the smallest number of assumptions is made on
  the WIMP velocity distribution in the halo of our Galaxy, it is
  possible to find values of the WIMP mass and the IDM mass
  splitting for which compatibility between present constraints and
  any of the three experiments claiming to see a WIMP excess among
  DAMA, CDMS-$Si$ and CRESST can be achieved.}

\begin{document}

\maketitle

\section{Introduction}
\label{sec:introduction}

Weakly Interacting Massive Particles (WIMPs) are among the best
motivated and most popular candidates to provide the Dark Matter that
is believed to constitute about 27\% of the total mass density of the
Universe, as confirmed by the latest measurement of cosmological
parameters\cite{planck}.  Presently, WIMPs are searched for by a
plethora of direct detection experiments, which look for the tiny
recoil energy $E_R$ imparted to the nuclei of a low--background
underground detector by the particles that are expected to constitute
the dark halo of our Galaxy. Due to the small expected $E_R$ (in the
keV range) and the tiny expected cross sections with ordinary nuclei
(10$^{-42}$ cm$^2$ or less in most popular scenarios) detecting WIMPs
is experimentally challenging. The present experimental situation is
quite elaborate, with two experiments claiming evidence for a yearly
modulation in their data attributable to a WIMP signal
(DAMA\cite{dama},CoGeNT\cite{cogent_modulation}), other two claiming a
non--observation in their modulation data
(CDMS\cite{cdms_ge_modulation} and KIMS\cite{kims_modulation}), some
others claiming a possibly WIMP--induced excess in their
time--averaged event spectra in tension with background estimates
(CoGeNT\cite{cogent_spectral}, CDMS-$Si$ \cite{cdms_si}, CRESST
\cite{cresst}) and many other experimental collaborations not
observing any discrepancy with the estimated background and as a
consequence publishing constraints that, when interpreted in specific
WIMP scenarios, are in (sometimes strong) tension with the
aforementioned results (LUX\cite{lux}, XENON100\cite{xenon100},
XENON10\cite{xenon10},KIMS\cite{kims}, CDMS-$Ge$\cite{cdms_ge},
CDMSlite \cite{cdms_lite}, SuperCDMS\cite{super_cdms}).

Until recently, the standard approach to interpret the results of
direct detection experiments in terms of a WIMP signal was to assume
for the WIMPs a specific velocity distribution in the Galactic
reference frame: in particular the usual assumption is the Isothermal
Sphere Model, i.e. a Maxwellian representing a WIMP gas in thermal
equilibrium with r.m.s. velocity $v_{rms}\simeq$ 270 km/sec and a
velocity upper cut representing the escape velocity. As suggested in
Ref.\cite{factorization}, however, when recoil energy intervals
analyzed by different experiments are mapped into same ranges for the
minimal velocity $v_{min}$ that the incoming WIMP needs to have to
deposit $E_R$ the halo model dependence can be factorized. Following
\cite{factorization} several analyzes have been performed to compare
the results of different experiments in terms of WIMP elastic
scattering without making specific assumptions on the WIMP velocity
distribution\cite{mccabe_eta,gondolo_eta}.

A scenario proposed to alleviate the tension among different direct
detection experiments is Inelastic Dark Matter (IDM)\cite{inelastic}.
In this class of models a Dark Matter (DM) particle $\chi$ of mass
$m_{DM}$ interacts with atomic nuclei exclusively by up--scattering to
a second state $\chi^{\prime}$ with mass
$m_{DM}^{\prime}=m_{DM}+\delta$.  In the case of exothermic Dark
Matter \cite{exothermic} $\delta<0$ is also possible: in this case the
particle $\chi$ is metastable and down--scatters to a lighter state
$\chi^{\prime}$. 

In both cases making use of the halo--model factorization approach is
significantly more complicated compared to the elastic case, because
when $\delta\ne$0 the mapping from $E_R$ to $v_{min}$ becomes more
involved than in the elastic case, introducing several complications:
for instance there is no longer a one--to--one correspondence between
$E_R$ and $v_{min}$.  So, while some early attempts exist
\cite{halo_independent_inelastic}, a systematic analysis of IDM where
an assessment of all the available data is done making use of the
factorization property of the halo--model dependence is still missing.
In the present paper we wish to address this issue, introducing some
strategies to determine regions in the IDM parameter space where the
tension existing among different experimental results can be (at least
partially) alleviated, and analyzing in detail some specific
benchmarks. In particular, by assuming a standard Maxwellian WIMP
velocity distribution, present data from
XENON100\cite{xenon100_inelastic} appear to already exclude an
interpretation of the DAMA modulation effect in terms of the IDM
hypothesis: as we will see, in specific cases the halo--dependence
factorization approach can indeed allow instead to find regions of the
IDM parameter space which are mutually compatible between DAMA and
liquid Xenon detectors.  However this approach can only be effective
when experimental results obtained using different detector targets
are compared. In fact, it is obvious that conflicting results obtained
with same--target detectors cannot be brought into agreement by any
theoretical assumption on the WIMP--nucleus scattering process,
including IDM. This is for instance the case for the apparent tension
between the DAMA\cite{dama} modulation result interpreted in terms of
WIMP--Iodine scatterings and the KIMS\cite{kims} claim of
non--observation of a WIMP excess with a $CsI$ target. Something
similar happens between the CoGeNT
excesses\cite{cogent_spectral,cogent_modulation} and upper bounds
obtained by other germanium detectors.  In this case the only way to
reconcile conflicting results is to look deeper in the possible
sources of systematic errors, including the many uncertainties
connected to quenching factors, atomic form factors, background cuts
efficiencies, etc.\cite{collar_liquid}.

In the following we chose to extend our analysis of IDM to
$\delta<0$. Notice however, that while this scenario can indeed
explain the excesses observed in the unmodulated spectra of direct
detection experiments (for instance it has been recently proposed to
explain the three WIMP--candidate events observed by CDMS-$Si$
\cite{exothermic_si}) in this regime DM is
expected to produce a very suppressed yearly modulation
signal\cite{exothermic}: in fact, when the kinetic energy of the
incoming WIMP is below the mass splitting $|\delta|$ the deposited
recoil energy $E_R$ is determined solely by the energy deposited in
the exothermic process, and is independent on the WIMP incoming
velocity. In this case if a yearly modulation is observed, it can
hardly be produced by the boost from the galactic to the Earth rest
frame.  As far as this aspect is concerned, we adopt a purely
phenomenological approach and chose to extend our analysis of the DAMA
modulation excess to $\delta<$0 without any theoretical
prejudice\footnote{In principle, a yearly modulation could still
  arise from some solar--system scale features in the DM spatial
  distribution.}.

The paper is organized as follows: in Section \ref{sec:factorization}
we summarize the halo--model factorization technique in WIMP--nucleus
scattering (also pointing out some of its limitations) and discuss the
problematics emerging in the context of IDM; in Section
\ref{sec:elastic} we show the present situation in the case of elastic
scattering; in Section \ref{sec:tests} we introduce some tests to
analyze the data which are specific to IDM; Section
\ref{sec:phenomenology_inelastic} contains the quantitative results of
this paper with a phenomenological discussion of the present
experimental results in the context of IDM; finally, in Appendix
\ref{app:exp} we summarize the experimental inputs used in the
analysis.

\section{Factorization of the halo model dependence}
\label{sec:factorization}

In the IDM scenario, in order to deposit the recoil energy $E_R$ in the
detector the incoming WIMP velocity $v$ (in the laboratory rest frame)
needs to be larger that the minimal speed $v_{min}$ :

\begin{equation}
v_{min}=\frac{1}{\sqrt{2 m_N E_R}}\left | \frac{m_NE_R}{\mu_{\chi N}}+\delta \right |,
\label{eq:vmin}
\end{equation}
\noindent where $m_N$ is the mass of the nucleus and $\mu_{\chi N}$ is
the WIMP--nucleus reduced mass.  Given a detector using a mono--atomic
target with active mass $M$ and exposition time $T$ the expected
differential rate for WIMP scatterings depositing the recoil energy
$E_R$ is given by:

\begin{equation}
\frac{dR}{dE_R}=M T \frac{\rho_{DM}}{m_{DM}} \sigma_0  \frac{N_T m_N\tilde{A}^2}{2 \mu_{\chi{\cal N}}^2} F^2(E_R) \eta(v_{min}),
\label{eq:diff_rate_er}
\end{equation}

\noindent where $\rho_{DM}$ is the DM mass density in the neighborhood
of the Sun, $N_T$ is the number of targets per unit mass, $\mu_{\chi
  {\cal N}}$ is the WIMP--nucleon (proton or neutron) reduced mass,
while:

\begin{equation}
  \sigma_0\equiv\lim_{v\rightarrow \infty} \sigma_p=\frac{\sigma_p}{\sqrt{1-\frac{2|\delta|}{\mu_{\chi N} v}}},
\end{equation}
\noindent with $\sigma_p$ the WIMP--proton point--like cross
section\footnote{In the IDM case $\sigma_p$ maintains a dependence on
  $v$ also in the non--relativistic limit, $\sigma_p\propto
  \sqrt{1-2|\delta|/(\mu_{\chi N} v)}$, but the dependence cancels out
  when calculating the differential cross section $d\sigma_p/d
  E_R$. For this reason the cross section factorized in
  Eq.(\ref{eq:diff_rate_er}) is $\sigma_0$ and not directly
  $\sigma_p$.}. Moreover $\tilde{A}$ is the ratio between the
WIMP--nucleus and the WIMP--proton interaction amplitudes.  In
particular, in the present paper we will assume the scaling law for a
scalar--coupling interaction:

\begin{equation}
\tilde{A}=Z+(A-Z)\frac{f_n}{f_p},
\label{eq:scaling_law}
\end{equation}
\noindent where $f_n/f_p$ represents the ratio between the coupling of
the $\chi$ particle to neutrons and protons, respectively, while $Z$
and $A$ are the atomic number and mass number of the target
nucleus. In Eq.(\ref{eq:diff_rate_er}) $F(E_R)$ is a form factor
taking into account the finite size of the nucleus for which we assume
the standard form\cite{helm}:

\begin{eqnarray}
F(E_R)&=&\frac{3}{qR^{\prime}}\left [ \frac{sin(qR^{\prime})}{(qR^{\prime})^2} - \frac{cos(qR^{\prime})}{qR^{\prime}}  \right]\exp\left (-\frac{(qs)^2}{2} \right )\\
q&=&\sqrt{2 m_N E_R};\;\; R^{\prime}=\sqrt{R_N^2-5 s^2};\;\;R_N=1.2 A^{\frac{1}{3}};\;\;s=\mbox{1 fm}.
\label{eq:form_factor}
\end{eqnarray}

\noindent Finally, in equation (\ref{eq:diff_rate_er}) the
function:

\begin{equation}
\eta(v_{min})=\int_{|\vec{v}|>v_{min}}\frac{f(\vec{v})}{|\vec{v}|}\; d^3 v,
\label{eq:eta}
\end{equation}

\noindent contains the dependence of the expected rate on the velocity
distribution $f(\vec{v})$ (boosted in the laboratory reference frame).

Eq. (\ref{eq:diff_rate_er}) can then be recast in the form:

\begin{equation}
\frac{dR}{dE_R}[E_R(v_{min})]=M T \frac{N_T m_N\tilde{A}^2}{2 \mu_{\chi{\cal N}}^2} F^2(E_R) \tilde{\eta}(v_{min}),
\label{eq:dr_de_recast}
\end{equation}

\noindent where the quantity:

\begin{equation}
\tilde{\eta}(v_{min})\equiv \frac{\rho_{DM}}{m_{DM}} \sigma_0 \eta(v_{min}),
\label{eq:eta_tilde}
\end{equation}

\noindent is a factor common to the WIMP--rate predictions of all
experiments, provided that it is sampled in the same intervals of
$v_{min}$. Mutual compatibility among different detectors' data can
then be investigated (factorizing out the dependence on the halo
velocity distribution) by binning all available data in the same set
of $v_{min}$ intervals and by comparing the ensuing estimations of
$\tilde{\eta}(v_{min})$.

In Eq.(\ref{eq:eta}) $f(\vec{v})$ represents the WIMP velocity
distribution boosted from the Galactic rest frame to the reference
frame of the laboratory. The involved boost depends on the velocity of
the Earth in the Galactic rest frame. Since the latter is the result
of the combination of the motion of the Solar system and of the
rotation of the Earth around the Sun a yearly time modulation is
predicted that can be used to discriminate between signal and
background in direct dark matter searches. In particular, in the case
of a Maxwellian WIMP velocity distribution this time dependence can be
approximated with the functional form:

\begin{equation}
\tilde{\eta}_{Maxwellian}(t) \simeq \tilde{\eta}_{Maxwellian,0}+\tilde{\eta}_{Maxwellian,1} \cos[\omega (t-t_0)],
\label{eq:cos_dependence}
\end{equation}

\noindent where $\omega=2\pi/365$ days, the phase $t_0\simeq$ 2 of
June corresponds to the moment when the velocity of the Sun in the
Galactic rest frame and the velocity of the Earth around the Sun point
in the same direction (leading to a maximum in the relative flux of
incoming WIMPS impinging on the detector) and the ratio
$\tilde{\eta}_{Maxwellian,1}/\tilde{\eta}_{Maxwellian,0}$ is predicted
to be between 5\% and 10\%.  Experiments sensitive to this yearly
modulation (such as DAMA and CoGeNT) actually provide estimations of
the modulated halo functions $\tilde{\eta_1}$ using the above
time--dependence, while the others get either estimations or upper
bounds on the unmodulated halo functions $\tilde{\eta}_0$.  Notice
however that, while the sinusoidal time dependence of the expected
rate given in Eq.(\ref{eq:cos_dependence}) is used by DAMA and CoGeNT
to analyze their modulation data, this is not the only possible one,
not even, for instance, in anisotropic extensions of the isothermal
sphere \cite{nic_ste}. Actually, strictly speaking the only halo model
independent definitions of the unmodulated and modulated parts of the
$\tilde{\eta}$ functions are simply:

\begin{eqnarray}
\tilde{\eta}_0 &\equiv&  <\tilde{\eta}(\Delta T=\mbox{1 year})>\nonumber \\
\tilde{\eta}_1 &\equiv&  \frac{<\tilde{\eta}(\Delta T_1)>-<\tilde{\eta}(\Delta T_2)>}{2},
\label{eq:eta_general}
\end{eqnarray}
\noindent where $<>$ represents time average, and $\Delta T_1$,
$\Delta T_2$ are two equal time intervals centered around the maximum
and minimum of the signal, whose phase should be determined directly
from the data and common to the analysis of the experiments that are
compared \footnote{Time averages of different experiments should be
  obtained using overlapping periods of data taking.}.

As a consequence of the above discussion, estimations of the
$\tilde{\eta}_1$ halo function which make use of annual modulation
amplitudes published by experimental collaborations are not really
halo independent, since their validity is restricted to the class of
halo models with the specific time dependence of
Eq.(\ref{eq:cos_dependence}). Nevertheless, in the following we will
make use of the published DAMA modulation amplitudes to estimate the
$\tilde{\eta}_1$ function, implicitly assuming that indeed the
$\tilde{\eta}_{Maxwellian,1}$ estimations obtained using the time
dependence (\ref{eq:cos_dependence}) do not differ significantly from
what one would obtain using Eq.(\ref{eq:eta_general}) to analyze the
data.

In a real--life experiment $E_R$ is obtained by measuring a related
detected energy $E^{\prime}$ obtained by calibrating the detector with
mono--energetic photons with known energy. However the detector
response to photons can be significantly different compared to the
same quantity for nuclear recoils.  For a given calibrating photon
energy the mean measured value of $E^{\prime}$ is usually referred to
as the electron--equivalent energy $E_{ee}$ and measured in keVee. On
the other hand $E_R$ (that represents the signal that would be
measured if the same amount of energy were deposited by a nuclear
recoil instead of a photon) is measured in keVnr. The two quantities
are related by a quenching factor $Q$ according to $E_{ee}=Q(E_R)
E_R$\footnote{The quenching factor is measured with a neutron source,
  and is subject to large uncertainties especially at low
  energies. Moreover, to avoid nuclear activation it is not measured
  in the same low--background detectors used for WIMP--search data.
  This is potentially a major source of systematics since $Q$ may
  depend on the operating experimental conditions and can vary in
  different detectors of the same material.}. Moreover the measured
$E^{\prime}$ is smeared out compared to $E_{ee}$ by the energy
resolution (a Gaussian smearing
$Gauss(E_{ee}|E^{\prime},\sigma_{rms}(E^{\prime}))\equiv
1/(\sigma_{rms}\sqrt{2\pi})exp[-(E^{\prime}-E_{ee})^2/(2\sigma_{rms}^2)]$
with standard deviation $\sigma_{rms}(E^{\prime})$ related to the Full
Width Half Maximum (FWHM) of the calibration peaks at $E^{\prime}$ by
$FHWM=2.35 \sigma_{rms}$ is usually assumed) and experimental count
rates depend also on the counting efficiency or cut acceptance
$\epsilon(E^{\prime})$.  Overall, the expected differential event rate
is given by:

\begin{equation}
  \frac{dR}{d E^{\prime}}=\epsilon(E^{\prime})\int_0^{\infty}d E_{ee} Gauss(E_{ee}|E^{\prime},\sigma_{rms}(E^{\prime}))\frac{1}{Q(E_R)} \frac{d R}{d E_R}.
\label{eq:rate_folding}
\end{equation}

In the case of liquid scintillators it is customary to measure both
the observed signal and its mean value directly in PE (photoelectrons)
and to parametrize the quenching factor in terms of an effective
light yield ${\cal L}_{eff}(E_R)$ normalized at the calibrating energy
of 122 keV. Moreover, lacking a direct calibration at low energy the
resolution is estimated assuming a Poisson fluctuation for the
photoelectrons and folding it with the Gaussian response of the
photomultiplier\cite{xenon_parametrization} (see Appendix
\ref{app:exp} for details).

Combining Eqs.(\ref{eq:diff_rate_er}) and (\ref{eq:rate_folding}) the
expected number of events in the interval
$E_1^{\prime}<E^{\prime}<E_2^{\prime}$ can be cast in the form:

\begin{equation}
  N_{theory}(E_1^{\prime},E_2^{\prime})=\int_{E_1^{\prime}}^{E_2^{\prime}} d E^{\prime} \frac{dR}{d E^{\prime}}=
  \int_{0}^{\infty} d E_{ee} \tilde{\eta}\left \{v_{min}\left [E_R\left (E_{ee} \right )  \right ] \right \} {\cal R}_{[E_1^{\prime},E_2^{\prime}]}(E_{ee}),
\end{equation}

\noindent where the response function ${\cal R}$, given by:

\begin{equation}
{\cal R}_{[E_1^{\prime},E_2^{\prime}]}(E_{ee})=\frac{N_T m_N \tilde{A}^2}{2 \mu_{\chi{\cal N}}^2}F^2\left[E_R(E_{ee}) \right ] MT 
\int_{E_1^{\prime}}^{E_2^{\prime}} d E^{\prime} Gauss(E_{ee}|E^{\prime},\sigma_{rms}(E^{\prime})) \epsilon(E^{\prime}), 
\label{eq:response_function}
\end{equation}

\noindent contains the information of each experimental
setup. Given an experiment with detected count rate $N_{exp}$ in the
energy interval $E_1^{\prime}<E^{\prime}<E_2^{\prime}$ the combination:

\begin{equation}
  \bar{\tilde{\eta}}=\frac{\int_{0}^{\infty} d E_{ee} \tilde{\eta}(E_{ee}) {\cal R}_{[E_1^{\prime},E_2^{\prime}]}(E_{ee})}
  {\int_{0}^{\infty} d E_{ee} {\cal R}_{[E_1^{\prime},E_2^{\prime}]}(E_{ee})}=\frac{N_{exp}}{\int_{0}^{\infty} d E_{ee} {\cal R}_{[E_1^{\prime},E_2^{\prime}]}(E_{ee})},
\label{eq:eta_bar_e}
\end{equation}

\noindent can be cast in the form\cite{gondolo_eta}:

\begin{equation}
  \bar{\tilde{\eta}}=\frac{\int_{0}^{\infty} d v_{min} \tilde{\eta}(v_{min}) {\cal R}_{[E_1^{\prime},E_2^{\prime}]}(v_{min})}
  {\int_{0}^{\infty} d v_{min} {\cal R}_{[E_1^{\prime},E_2^{\prime}]}(v_{min})}=\frac{N_{exp}}{\int_{0}^{\infty} d v_{min} {\cal R}_{[E_1^{\prime},E_2^{\prime}]}(v_{min})},
\label{eq:eta_bar_vmin}
\end{equation}

\noindent by changing variable from $E_{ee}$ to $v_{min}$ (in the
above expression ${\cal R}_{[E_1^{\prime},E_2^{\prime}]}(v_{min})$ =
${\cal R}_{[E_1^{\prime},E_2^{\prime}]}(E_{ee})$ $d E_{ee}/d v_{min}$)
and can be interpreted as an average of the function
$\tilde{\eta}(v_{min})$ in an interval
$v_{min,1}<v_{min}<v_{min,2}$. The latter is defined as the one where
the response function ${\cal R}$ is ``sizeably'' different from zero
(we will conventionally take the interval
$v_{min}[E_R(E_{ee,1})]<v_{min}<v_{min}[E_R(E_{ee,2})]$ with
$E_{ee,1}=E^{\prime}_1-\sigma_{rms}(E^{\prime}_1)$,
$E_{ee,2}=E^{\prime}_2+\sigma_{rms}(E^{\prime}_2)$, i.e. the
$E^{\prime}$ interval enlarged by the energy resolution).

The formalism summarized in this Section is no longer straightforward
in the case of experiments that use multi--atomic targets, unless it
is possible to conclude that only one of them dominates the
scattering\footnote{The case of two targets very close in mass is one
  exception: see Appendix \protect\ref{app:exp} for a discussion of
  $CsI$ in KIMS\cite{kims}.}. This is the case of NaI in DAMA, where
for the elastic case it is possible to conclude that the $v_{min}$
range required to explain scatterings off the Iodine nuclei is above
reasonable values of the escape velocity for $m_{DM}\lsim$ 20 GeV, so
that at low WIMP masses domination of scatterings off Na nuclei can be
established; on the other hand at larger masses, assuming the scaling
law in (\ref{eq:scaling_law}) with $f_n/f_p=1$, scatterings on Iodine
are automatically guaranteed to be predominant (see
Fig.\ref{fig:scaling_laws} and discussion in Section
\ref{sec:elastic}). However, in the case of inelastic scattering the
situation is more involved, as will be discussed at the end of Section
\ref{sec:phenomenology_inelastic}.

The factorization procedure described above can be separately applied
in a straightforward way to get estimates $\bar{\tilde{\eta}}_0$ and
$\bar{\tilde{\eta}}_1$ of the constant and modulated parts of the
$\bar{\tilde{\eta}}$ function. As already pointed out before, however,
we wish to remark again that in the case of the modulated part
experimental amplitudes are usually extracted by assuming the time
dependence of Eq.(\ref{eq:cos_dependence}), limiting {\it de facto}
the scope of a ``halo--independent'' analysis to a restricted (albeit
well motivated) class of models.

\section{Phenomenological analysis: the case of elastic scattering}
\label{sec:elastic}

\begin{figure}[h]
\begin{center}
\includegraphics[width=0.7\columnwidth,bb= 46 194 506 635]{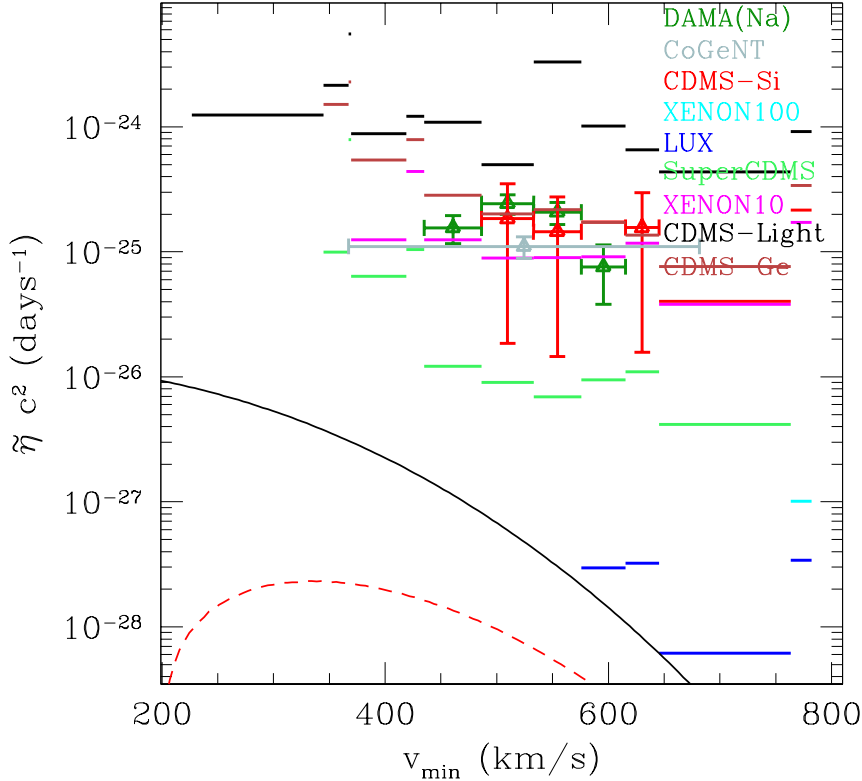}
\end{center}
\caption{Measurements and bounds for the functions
  $\tilde{\eta}_0$,$\tilde{\eta}_1$ defined in
  Eq.(\protect\ref{eq:eta_general}) for $m_{DM}$=8 GeV, $\delta$=0
  (elastic scattering) and $f_p=f_n$=1. Open triangles represent
  excesses that can be interpreted as WIMP signal candidates (the
  vertical error bars are at 1 $\sigma$) while horizontal lines
  represent 90\% C.L. upper bounds from experiments that did not
  observe any excess. The solid (black) and dashed (red) lines
  represent the predictions for $\tilde{\eta}_0$ and $\tilde{\eta}_1$,
  respectively, for a Maxwellian velocity distribution, normalized to
  the most constraining upper bound.}
\label{fig:benchmark_mchi_8_delta_0}
\end{figure}

\begin{figure}[h]
\begin{center}
\includegraphics[width=0.7\columnwidth,bb= 46 194 506 635]{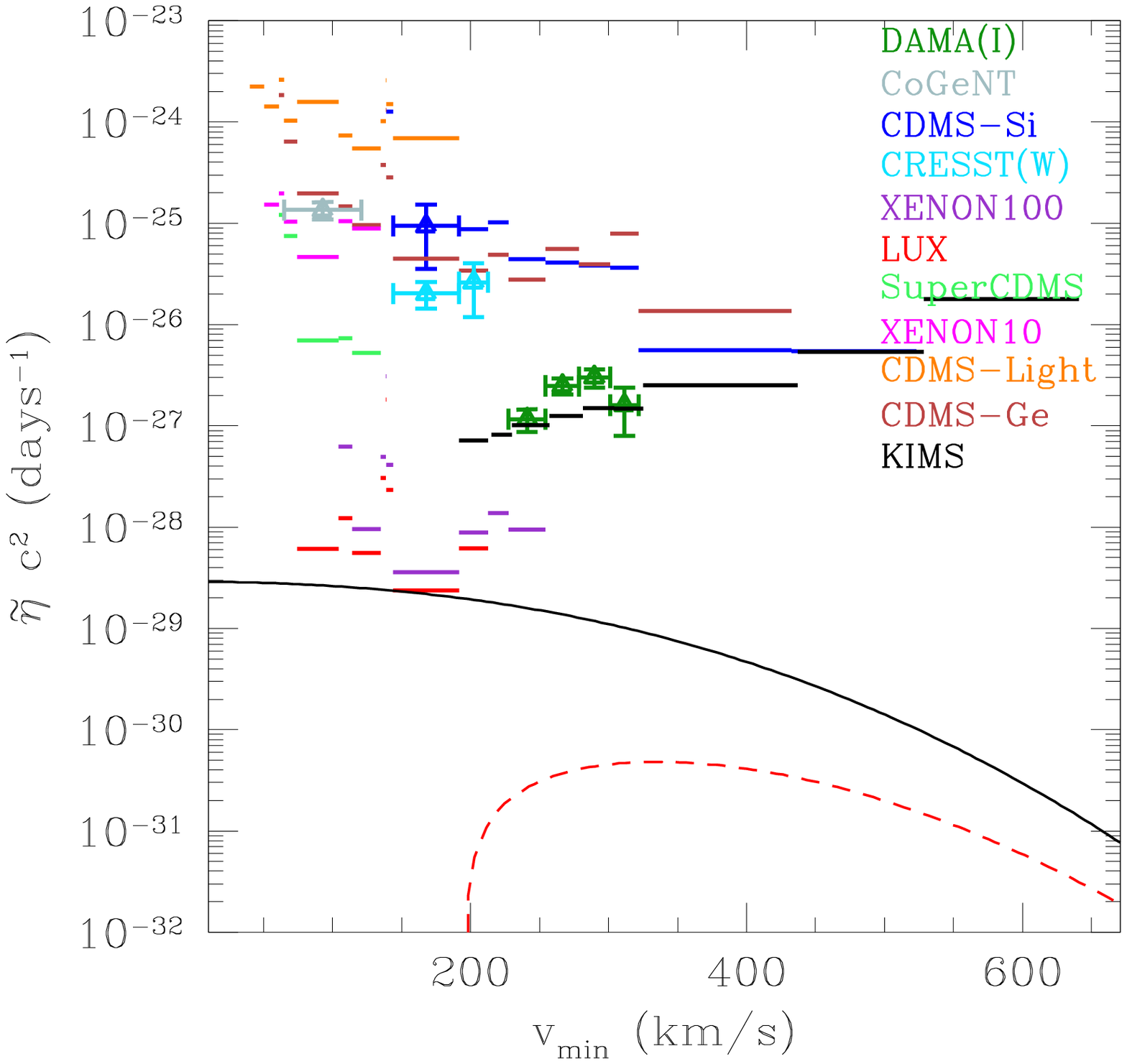}
\end{center}
\caption{The same as in Fig.\ref{fig:benchmark_mchi_8_delta_0} with $m_{DM}$=100GeV.}
\label{fig:benchmark_mchi_100_delta_0}
\end{figure}

In Figures \ref{fig:benchmark_mchi_8_delta_0} and
\ref{fig:benchmark_mchi_100_delta_0} we summarize the present
experimental situation for the determination of the
$\bar{\tilde{\eta}}_0$ and $\bar{\tilde{\eta}}_1$ functions in the case
of elastic scattering ($\delta=0$) as a function of $v_{min}$ for the
two representative values $m_{DM}$=8 GeV and $m_{DM}$=100 GeV.  In
both figures we adopt $f_n/f_p$=1, assuming in DAMA scatterings on Na when
$m_{DM}$=8 GeV and scattering on I when $m_{DM}$=100 GeV.

We include in the analysis the following experiments: DAMA\cite{dama},
CoGeNT \cite{cogent_spectral}, CDMS-$Si$\cite{cdms_si},
XENON100\cite{xenon100},LUX \cite{lux} SuperCDMS \cite{super_cdms}
XENON10 \cite{xenon10}, CDMSlite \cite{cdms_lite}, CDMS-$Ge$
\cite{cdms_ge}, CRESST\cite{cresst}, KIMS\cite{kims}. The details of
the parameters used to evaluate the response function of each
experiment are summarized in Appendix \ref{app:exp}.

In this Section and in the following ones we choose to map all the
experimental data in the same $v_{min}$ intervals, in order to
directly compare different determinations of the same quantities
$\bar{\tilde{\eta}}_{0,1}$ averaged within identical ranges of
$v_{min}$.  Since DAMA is the experiment with the highest accumulated
statistics and the strongest indication of a possible WIMP signal we
rebin the data of all other experiments in the $v_{min}$ intervals
corresponding to the energy bins of DAMA. Outside the $v_{min}$ range
covered by DAMA we bin the data using the collection of $v_{min}$
values obtained considering for each experiment the energy boundary of
the corresponding Region Of Interest(ROE)=[$E_{min},E_{max}$] and
collecting the two $v_{min}$ values corresponding to $E_{min}$ and
$E_{max}$. This automatic binning procedure implies that in
Figs.\ref{fig:benchmark_mchi_8_delta_0} and
\ref{fig:benchmark_mchi_100_delta_0} of this Section and in the
analogous figures in the following ones some velocity bins may happen
to be small, so that, depending on the resolution of the plots, some
of the $\bar{\tilde{\eta}}_{0,1}$ determinations or upper bounds may
appear as dots.

In agreement with previous analyzes\cite{gondolo_eta}, taken at face
value the results of Figs. \ref{fig:benchmark_mchi_8_delta_0} and
\ref{fig:benchmark_mchi_100_delta_0} show a strong tension among
different results. In particular, the null results of SuperCDMS, LUX
and XENON100 imply constraints on $\bar{\tilde{\eta}}_0$ which are
about one and two orders of magnitude smaller compared to the
$\bar{\tilde{\eta}}_1$ ranges suggested by DAMA, or the
$\bar{\tilde{\eta}}_0$ ranges suggested by CDMS-$Si$, CRESST or
CoGeNT\footnote{It should be noted that the LUX and XENON100 bounds
  for $m_{DM}$=8 GeV are obtained by using the low--energy part of the
  spectrum ($S_1<$ 3 PE), where the sensitivity of liquid
  scintillators has been put into question
  \cite{collar_liquid}. However also the SuperCDMS and XENON10 data
  appear constraining , albeit to a somewhat less extent. In
  particular, the energy range determining the SuperCDMS bound is
  $E_R<$7 keV while the corresponding one for XENON10 is $E_R<$4.3
  keV.  See Appendix \ref{app:exp} for details on various
  experiments.}. Moreover, for $m_{DM}=8$ GeV there also appears to be
tension among the DAMA and CDMS-$Si$ excesses, with the
$\bar{\tilde{\eta}}_1$ ranges indicated by DAMA in the upper range of
the $\bar{\tilde{\eta}}_0$ values suggested by CDMS-$Si$. Finally, in
Fig. \ref{fig:benchmark_mchi_100_delta_0}, where the DAMA excess is
explained by WIMP--$I$ scatterings, the KIMS upper bound is in tension
with the DAMA modulation data. As already anticipated in the
Introduction, this discrepancy cannot be alleviated by assuming a
different scenario for the WIMP--nucleus interaction and will persist
in the IDM case discussed in Section \ref{sec:phenomenology_inelastic}
whenever WIMP--$I$ scatterings are assumed in DAMA. The same happens
for the apparent discrepancy between the CoGeNT excess and the
SuperCDMS bound, which both use Germanium targets.

The tension discussed above is even stronger if a Maxwellian is
assumed for the velocity distribution: in both
Figs. \ref{fig:benchmark_mchi_8_delta_0} and
\ref{fig:benchmark_mchi_100_delta_0} the solid and dashed lines
represent the corresponding predictions for $\tilde{\eta}_0$ and
$\tilde{\eta}_1$, respectively, normalized to the most constraining
limit. In this case, for instance, the DAMA result stands between two
and three orders of magnitude above the maximal value of
$\tilde{\eta}_1$ compatible to the LUX upper bound. In the following
Sections we will discuss possible strategies to relieve (at least
partially) the observed discrepancies discussed above within the
context of IDM and if the smallest possible number of theoretical
assumptions is made on the functions $\tilde{\eta}_0$ and
$\tilde{\eta}_1$.

\section{Halo--independent tests for Inelastic Dark Matter}
\label{sec:tests}

A complication of the IDM case compared to elastic scattering is that
the mapping between $v_{min}$ and $E_R$ (and so $E^{\prime}$) from
Eq. (\ref{eq:vmin}) is no longer univocal. In particular 
$v_{min}$ has a minimum when $E_R$=$E_R^*$=$|\delta|\mu_{\chi N}/m_N$ given by:

\begin{equation}
v_{min}^*=\left \{ \begin{array}{ll}
\sqrt{\frac{2|\delta|}{\mu_{\chi N}}} & \mbox{if $\delta>0$ }\\
0 &   \mbox{if $\delta<0$},
\end{array}
  \right . 
\label{eq:vstar}
\end{equation}

\noindent and any interval of $v_{min}>v_{min}^*$ corresponds to two
mirror intervals for $E_R$ with $E_R<E_R^*$ or $E_R>E_R^*$.  As a
consequence of this when $E_{ee}(E^*_R)\in [E_{ee,1},E_{ee,2}]$ the
change of variable from Eq.(\ref{eq:eta_bar_e}) to
Eq.(\ref{eq:eta_bar_vmin}) leads to two disconnected integration
ranges for $v_{min}$ and to an expression of $N_{theory}$ in terms of
a linear combination of the corresponding two determinations of
$\bar{\tilde{\eta}}$. This problem can be easily solved by binning the
energy intervals in such a way that for each experiment the energy
corresponding to $E_{ee}(E_R^*)$ is one of the bin boundaries.  So we
generalize the $v_{min}$ binning procedure described in Section
\ref{sec:elastic}: first we rebin (if needed) the DAMA data, starting
from $E_{ee}(E_R^*)$ and selecting ``mirror'' energy intervals lower
and higher than $E_{ee}(E_R^*)$ in such a way that they correspond to
equal $v_{min}$ ranges (we adopt the requirement that the smallest of
the two mirror bins so obtained is equal to 0.5 keV and discard
smaller values).  Outside the $v_{min}$ range covered by DAMA we bin
the data using the collection of $v_{min}$ values obtained considering
for each experiment the energy boundary of the corresponding Region Of
Interest=[$E_{min},E_{max}$] and collecting the three (two) $v_{min}$
values corresponding to $v_{min}^*$ (when applicable) and to
$E_{min}$, $E_{max}$. In the case $\delta \ne 0$ this procedure
ensures that $E_{ee}(E^*_R)\notin [E_{ee,1},E_{ee,2}]$ for all the
shown data. Provided that $E_{ee}(E^*_R)\notin [E_{ee,1},E_{ee,2}]$
the procedure to determine the $\bar{\tilde{\eta}}$ averages is then
similar to the elastic case.

In particular the change of variable from $E_{ee}$ to $v_{min}$
leading to Eq. (\ref{eq:eta_bar_vmin}) depends now on the $\delta$
parameter, so that the response function ${\cal R}$ not only depends
on the target mass $m_N$ but also on $\delta$. However, it is clear
that given an energy interval and a corresponding experimental count
rate the denominator of Eq.(\ref{eq:eta_bar_e}) does not depend on
$\delta$. The only effect of a change of $\delta$ is then a shift in
the corresponding $v_{min}$ range.

We notice here that, with the exception of DAMA and KIMS, all the
experiments that we will discuss in the following have made their
count rates public so that rebinning their data according to the
required $v_{min}$ ranges will be straightforward. In the case of the
DAMA data, whenever possible we will make direct use of the binned
modulated amplitudes published by the Collaboration\cite{dama} and map
all the other experimental results in those bins. Notice however that,
as explained above, we will need to rebin also the DAMA data whenever,
on Sodium or Iodine, $E_{ee}(E^*_R)\in [E_{ee,1},E_{ee,2}]$. This
would require to have the raw count rates. However, given the large
statistics of the DAMA data we decide to proceed by averaging the
binned modulated amplitudes $S_{m,k}$ taken from \cite{dama} in the
new energy bins using the expression:

\begin{equation}
  S_m^{rebinned}(E_{min},E_{max})=\frac{1}{\Delta E} \sum_{k} \bar{\Delta} E_k S_{m,k},
\label{eq:dama_rebin}
\end{equation}

\noindent where $\Delta E=E_{min}$--$E_{max}$ is the width of the new
bin and $\bar{\Delta} E_i$ is the overlap between $\Delta E$ and the
original bin $\Delta E_i=E_{i+1}-E_i$. To get an estimate of the
fluctuation on $S_m^{rebinned}$ we conservatively use the above
formula where the $S_{m,i}$ are replaced by their corresponding
1--$\sigma$ upper and lower values as taken from \cite{dama}. We will
adopt the same procedure also when rebinning the KIMS data.

For later convenience, let us now introduce some notation that will be
useful in the following Sections.  We will follow the convention of
naming $v_{min}$ intervals using capital--lettered names starting with
V (for instance,
\verb+V_DAMA_NA+$\equiv[v_1^{DAMA,Na},v_2^{DAMA,Na}]$) and the
corresponding energy intervals with the same name where the initial V
is substituted by an E, using an arrow to indicate the mapping of one
into the other. Moreover, since for $\delta\ne$0 each energy interval
has a mirror one corresponding to the same $v_{min}$ range we will add
an initial \verb+M_+ to the name of an energy interval to indicate
it. So: \verb+V_DAMA_NA+$\rightarrow$
\verb+E_DAMA_NA,M_E_DAMA_NA+\footnote{If the mirror interval of the
  signal range does not belong to the ROE a one--to--one
  correspondence between $v_{min}$ and the recoil energy is
  recovered.}. Moreover we will combine intervals using logical and
simple mathematical symbols (for instance, V1$\cap$V2$>$V3 means that
the $v_{min}$ values belonging to the overlapping of V1 and V2 are all
larger than those belonging to V3).
  
In the present analysis we wish to make the smallest possible number
of assumptions on the two functions $\tilde{\eta}_0$ and
$\tilde{\eta}_1$. In particular, they reduce to:

\begin{eqnarray}
&\tilde{\eta}_0(v_{min,2})\le\tilde{\eta}_0(v_{min,1})  & \mbox{if $v_{min,2}> v_{min,1}$},\nonumber\\ 
& \tilde{\eta}_1\le\tilde{\eta}_0  & \mbox{at the same $v_{min}$},\nonumber\\
& \tilde{\eta}_0(v_{min} \ge v_{esc})=0. & 
\label{eq:eta_conditions}
\end{eqnarray}

The first condition descends from the definition (\ref{eq:eta}), that
implies that $\tilde{\eta}(v_{min})$ is a decreasing function of
$v_{min}$. The second is an obvious consequence of the fact that
$\tilde{\eta}_1$ is the modulated part of $\tilde{\eta}$. The last
condition reflects the requirement that the WIMPs are gravitationally
bound to our Galaxy. In the following we will assume that the WIMP
halo is at rest in the Galactic rest frame and we will adopt as the
maximal velocity of WIMPs $v_{esc}$=782 km/s in the reference frame of
the laboratory, by combining the reference value of the escape
velocity $v_{esc}^{Galaxy}$=550 km/s in the galactic rest frame with
the velocity $v_0$=232 km/s of the Solar system with respect to the
WIMP halo. Note that the specific value of the escape velocity is
relevant at low values of $m_{DM}$, for which positive excesses can be
explained by ranges of $v_{min}$ close to $v_{esc}$.  For each of the
benchmarks discussed in the following the relevance of $v_{esc}$ will
be easily read--off from the corresponding
$v_{min}$-$\bar{\tilde{\eta}}_{0,1}$ plot.

\subsection{Internal consistency checks}
\label{sec:shape_test}

We describe here two internal checks (i.e. involving the data of one
single experiment) that we will apply in a systematic way in the
discussion of Section \ref{sec:phenomenology_inelastic}.

Suppose that a direct detection experiment observes an excess
potentially attributable to a WIMP signal in the energy range
\verb+E_SIG+ (this will require the additional condition \verb+E_SIG+
$\rightarrow$ \verb+V_SIG+ $\in$
\verb+V_GAL+$\equiv[0,v_{esc}]$). Then, according to the discussion
below Eq. (\ref{eq:vstar}), if E$^*\in $ \verb+E_SIG+ (where $E^*$ is
defined above Eq.(\ref{eq:vstar})) the interval \verb+E_SIG+ must be
split into two sub--intervals \verb+E_SIG1+ and \verb+E_SIG2+, with
\verb+E_SIG1+$\rightarrow [v_{min}^*,v_{min}^1]\equiv$\verb+V_STAR1+
and \verb+E_SIG2+$\rightarrow
[v_{min}^*,v_{min}^2]\equiv$\verb+V_STAR2+. Then, if for instance
\verb+V_STAR1+ $\subset $ \verb+V_STAR2+ one has \verb+M_E_SIG1+
$\subset $ \verb+E_SIG2+, and
\verb+E_SIG1+,\verb+M_E_SIG1+$\rightarrow$ \verb+V_STAR1+, i.e. the
two intervals allow for independent determinations of the function
$\tilde{\eta}_0$ or $\tilde{\eta}_1$ in the same $v_{min}$ interval
(by the same token if instead \verb+V_STAR2+ $\subset $ \verb+V_STAR1+
then \verb+E_SIG2+,\verb+M_E_SIG2+$\rightarrow$ \verb+V_STAR2+). This
procedure allows to perform a compatibility check, as already pointed
out in \cite{halo_independent_inelastic}. In the following we will
refer to this procedure as the ``shape test''.

\begin{figure}[h]
\begin{center}
\includegraphics[width=0.7\columnwidth]{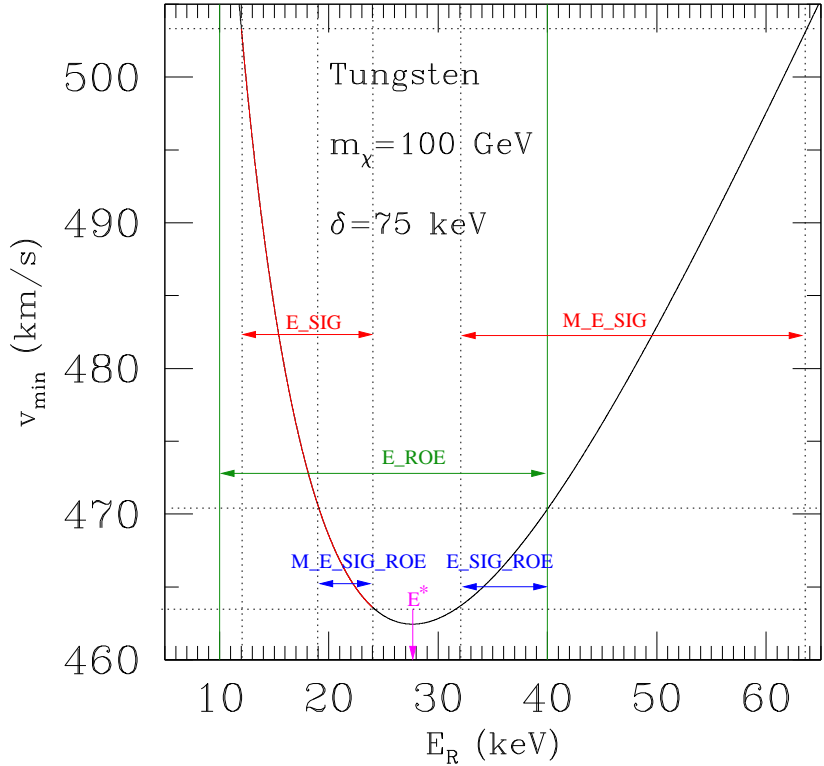}
\end{center}
\caption{Schematic view of the mirror test introduced in Section
  \protect\ref{sec:tests}, for $m_{DM}$=100 GeV, $\delta$=75 keV in
  the specific case of Tungsten in CRESST. The red part of the
  parabolic curve corresponds to the energy range [12 keV,24 keV] of
  the CRESST excess, while the interval $E_R$=[10 keV,40 keV]
  corresponds to the analyzed Region of Interest (see Appendix
  \protect\ref{app:exp}).}
\label{fig:mirror_test}
\end{figure}

Notice that the shape test can be effective only in the (small) range
of the $m_{DM}$ and $\delta$ parameters where E$^*\in$ \verb+E_SIG+.
Actually, a more general and potentially more constraining test can be
devised when E$^*\notin$ \verb+E_SIG+ and the mirror interval
\verb+M_E_SIG+ of \verb+E_SIG+ corresponds to an energy range where no
signal has been detected. Let us indicate with \verb+E_ROE+ the
complete energy interval analyzed by the experiment. Then, the two
mirror intervals \verb+E_SIG_ROE+$ \equiv$ \verb+M_E_SIG+ $\cap$
\verb+E_ROE+ and \verb+M_E_SIG_ROE+ $\in$ \verb+E_SIG+ correspond to
the same $v_{min}$ interval and allow to perform a compatibility test
analogous to the shape test. In the following we apply this procedure
referring to it as the ``mirror test''. For the sake of clarity, we
schematically outline this procedure in Fig. \ref{fig:mirror_test} for
$m_{DM}$=100 GeV, $\delta$=75 keV in the specific case of Tungsten in
CRESST, where (see Appendix \ref{app:exp}) \verb+E_ROE+=[10 keV,40
keV] and \verb+E_SIG+=[12 keV,24 keV].

We notice here that, on general grounds, taking experimental energy
bins smaller than the energy resolution is questionable. For this
reason in order to perform either the shape or the mirror test
described above we will require that the widths of the two
energy intervals involved are larger than the energy resolution.

In the case of DAMA, due to the large statistics we will assume that
the two independent determinations of $\bar{\tilde{\eta}}_{1}$,
indicated with $\bar{\tilde{\eta}}_{1,1}$ and
$\bar{\tilde{\eta}}_{1,2}$, have Gaussian fluctuations. Denoting with
$\sigma_1$ and $\sigma_2$ the corresponding statistical errors, we
will then require, as in \cite{halo_independent_inelastic}:

\begin{equation}
\Delta_{ST}\equiv \frac{\left |\bar{\tilde{\eta}}_{1,1}-\bar{\tilde{\eta}}_{1,2}\right |}{\sqrt{\sigma_1^2+\sigma_2^2}}\le 1.64,
\label{eq:shape_test}
\end{equation}
\noindent at the 95 \% C.L.

The other experiments claiming some excess that we will discuss in the
following (CRESST and CDMS-$Si$) are not sensitive to the annual
modulation but only to absolute (time--averaged) rates; moreover their
statistics is much reduced compared to DAMA. In this case we assume
Poissonian fluctuations when comparing the two mirror energy bins. Let
us indicate them with \verb+E_SIG+ and \verb+M_E_SIG+.  In order to
check if two independent count rates $N_1$ and $N_2$ belong to two
Poisson distributions with averages $\lambda_1$ and $\lambda_2$ the
conditional test by Przyborowski and Wilenski can be
adopted\cite{poisson_test}. In particular the p--value for the
hypothesis $\lambda_1/\lambda_2\le c$ is given by:

\begin{equation}
p=\sum_{i=N_1}^{N_1+N_2}\left ( \begin{array}{c}N_1+N_2 \\ i \end{array} \right )
[f(c)]^i [1-f(c) ]^{N_1+N_2-i},
\label{eq:p_value_poisson}
\end{equation}

\noindent with $f(c)=c/(1+c)$. At 95\% C.L. we will then require $p\ge$0.05 with:

\begin{equation}
\frac{\lambda_1}{\lambda_2}\le c=\frac{\int_{0}^{\infty} d E_{ee} {\cal R}_{\mbox{E} 
\raisebox{-0.5ex}{-}
\mbox{SIG}}(E_{ee})}
{\int_{0}^{\infty} d E_{ee} {\cal R}_{\mbox{M} 
\raisebox{-0.5ex}{-}\mbox{E} 
\raisebox{-0.5ex}{-}
\mbox{SIG}}(E_{ee})}.
\end{equation}
\noindent

We conclude this Section by noting that a different type of internal
check is needed in the case of multi--target experiments, such as
DAMA and CRESST. As pointed out in Section \ref{sec:factorization} in
this case the factorization of the halo model dependence is only
possible if dominance of scatterings off one single target can be
established.  Consider for instance $NaI$ in DAMA: the energy range of
the signal is mapped into two different $v_{min}$ ranges corresponding
to scatterings off $Na$ and $I$, i.e.: \verb!E_SIG!  $\rightarrow$
\verb!V_SIG_NA!, \verb!V_SIG_I!. When $m_{DM}$ is small enough one has
dominance of scatterings off $Na$ since \verb!V_SIG_NA!$\subset$
\verb!V_GAL! and \verb!V_SIG_I!$\nsubseteq$ \verb!V_GAL!. On the other
hand at higher values of the WIMP mass the situation occurs when
\verb!V_SIG_NA!, \verb!V_SIG_I!$\subset$ \verb!V_GAL! with
\verb!V_SIG_I!$>$ \verb!V_SIG_NA!. In the standard situation where a
Maxwellian velocity distribution and a scalar coupling with
$f_n/f_p$=1 are assumed one has $\tilde{\eta}_1$(\verb!V_SIG_I!)$\ll
\tilde{\eta}_1$(\verb!V_SIG_NA!) but the (small) ratio
$\tilde{\eta}_1$(\verb!V_SIG_I!)$/ \tilde{\eta}_1$(\verb!V_SIG_NA!) is
fixed in the Maxwellian case, so that the scaling law in
(\ref{eq:scaling_law}) overcomes it in favour of dominance of
scatterings off Iodine. Notice, however, that if no assumptions are
made on the functional form of $\tilde{\eta}_1$ it is in principle
possible to assume that whenever \verb!V_SIG_I!$\cap$
\verb!V_SIG_NA!=0 either $\tilde{\eta}_1$(\verb!V_SIG_I!)  or
$\tilde{\eta}_1$(\verb!V_SIG_NA!)  is small enough to allow for
dominance of the other target.  Notice that $\tilde{\eta}_1$ is not
required to be decreasing with $v_{min}$ and can have in principle any
functional form.

The situation of multi--target experiments observing an excess in
$\tilde{\eta}_0$ is slightly more constrained, due to the requirement
that $\tilde{\eta}_0$ is decreasing with $v_{min}$. Given two targets
\verb!E_SIG!  $\rightarrow$ \verb!V_SIG_TARGET_1!,
\verb!V_SIG_TARGET_2!  and whenever the scaling law favours
\verb!TARGET_2!, dominance of \verb!TARGET_1! is realized if
$\tilde{\eta}_0$(\verb!V_SIG_TARGET_2!)$\simeq$0. However this
requires \verb!V_SIG_TARGET_2! $>$ \verb!V_SIG_TARGET_1! strictly. Of
course a particular case of this is the stronger requirement
\verb!V_SIG_TARGET_2!$>v_{esc}$ (notice that for inelastic scattering
this does not automatically mean that \verb!TARGET_2! is the heavier
target, since, when $\delta\ne$0, $v_{min}$ in Eq. (\ref{eq:vmin}) is
no longer necessarily increasing with the target mass).

\subsection{Comparison among different experiments}
\label{sec:vmin_regions}

As discussed in Section \ref{sec:elastic} for the elastic case,
present DM data show tension between excesses on $\tilde{\eta}_0$ and
$\tilde{\eta}_1$ and constraints on $\tilde{\eta}_0$, and the extent
of the discrepancy is particularly strong if an Isothermal Sphere
model for the WIMP velocity distribution is assumed.  In particular,
inspection of Figs. \ref{fig:benchmark_mchi_8_delta_0} and
\ref{fig:benchmark_mchi_100_delta_0} shows that, when no model is
assumed for the velocity distribution, an experiment can constrain
another one if it is sensitive to the same $v_{min}$ interval, or to
{\it lower} values (the latter condition descending from the
requirement that $\tilde{\eta}_0$ is decreasing monotonically with
$v_{min}$). However, both the condition that $v_{min}<v_{esc}$ in the
range explaining a possible excess, and the degree of overlapping
between two experiments using different target materials depend on the
mapping between $E_R$ and $v_{min}$, which, according to
Eq.(\ref{eq:vmin}), in the IDM scenario can be modified by assuming
$\delta\ne$0. In particular, if for the same choice of $m_{DM}$ and
$\delta$ conflicting experimental results can be mapped into
non--overlapping ranges of $v_{min}$ and if the $v_{min}$ range of the
constraint is at higher values compared to the excess (while that of
the signal remains below $v_{esc}$) the tension between the two
results can be eliminated by an appropriate choice of the
$\tilde{\eta}_{0,1}$ functions in compliance with the conditions of
Eq. (\ref{eq:eta_conditions}). This, of course, at the price of having
to assume that $\tilde{\eta}_0$ and $\tilde{\eta}_1$ drop to
appropriately low values in the (high) $v_{min}$ range pertaining to
the constraint.

The requirements expressed above can be expressed in a compact form by
making use of the notation introduced at the beginning of this
Section.  Suppose that one experiment measures an excess over the
background in the energy interval \verb!E_SIG! while the result of the
search of a second experiment is null in the energy range
\verb!E_BOUND!. Then to have compatibility between the signal
interpretation and the experimental bound, given
\verb!E_SIG!$\rightarrow$\verb!V_SIG!,
\verb!E_BOUND!$\rightarrow$\verb!V_BOUND!, one needs to impose the two
conditions: \verb!V_SIG!$\subset$\verb!V_GAL! and
\verb!V_BOUND!$>$\verb!V_SIG!.

\section{Phenomenological analysis: the case of inelastic scattering}
\label{sec:phenomenology_inelastic}

In this Section we wish to extend the results of Section
\ref{sec:elastic} to the inelastic case, by making use of the criteria
introduced in Section \ref{sec:tests}.  In order to proceed, we first
notice that, as explained in Section \ref{sec:factorization}, the
effect of mapping energies to velocities using Eq.(\ref{eq:vmin}) with
$\delta\ne$0 is only to shift the $\bar{\tilde{\eta}}_{1,2}$
determinations in the $v_{min}$ space compared to the $\delta$=0 case,
without changing their normalization (with the exception of moderate
changes due to possible modifications in the data binning). As a
consequence of this, we can conclude that, at least qualitatively, we
expect the hierarchy in the impacts of different constraints as
observed in Figs.\ref{fig:benchmark_mchi_8_delta_0} and
\ref{fig:benchmark_mchi_100_delta_0} to be preserved in the case of
inelastic scattering, with XENON100, LUX and the SuperCDMS remaining
the most constraining bounds, in particular between one and three
orders of magnitude below the DAMA result. So our approach will be to
first explore systematically the $m_{DM}$--$\delta$ parameter space to
find regions where the XENON100, LUX and the SuperCDMS constraints are
relaxed, and then to pick within those regions some representative
benchmark points where to discuss in more detail the experimental
situation including all the other bounds.

On the other hand, notice that, as already pointed out in our
Introduction, while the discrepancy between KIMS and the DAMA
interpretation in terms of WIMP--$I$ scatterings appears to be
quantitatively less severe, in this case it cannot be relieved by
assuming the IDM scenario. Something similar happens between the
CoGeNT effect and other bounds obtained with Germanium detectors. For
this reason we will include the CoGeNT effect and the KIMS bound in
all the relevant plots, but we will not consider them in our
discussion of compatibility ranges in the $m_{DM}$--$\delta$ parameter
space.

In the following we will analyze the IDM parameter space using the
parameter ranges: 

\begin{equation}
1\; \mbox{GeV} \le m_{DM} \le \mbox{1 TeV}, -300\; \mbox{keV} \le \delta \le 300\; \mbox{keV}.
\label{eq:mchi_delta_ranges}
\end{equation}

\subsection{Sodium scattering in DAMA and the CDMS-$Si$ excess}
\label{sec:dama_na_low_mass}

\begin{figure}[h]
\begin{center}
\includegraphics[width=0.7\columnwidth,bb= 46 194 506 635]{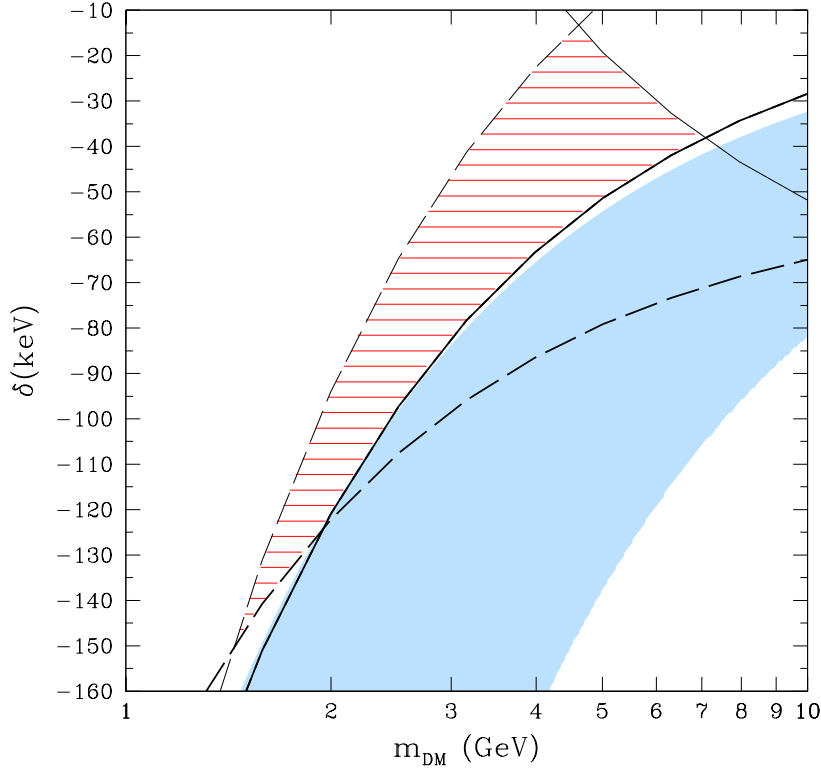}
\end{center}
\caption{Mass splitting $\delta=m^{\prime}_{DM}-m_{DM}$ as a function
  of $m_{DM}$.  The horizontally (red) hatched area represents the IDM
  parameter space where the modulation effect measured by DAMA
  assuming scattering on Sodium corresponds to a $v_{min}<v_{esc}$
  range which is always below the corresponding one probed by LUX and
  XENON100. As explained in the text, in this case Xenon experiments
  can constrain the DAMA excess only when some assumptions are made on
  the galactic velocity distribution. The enclosed region is the
  result of the combination four boundaries (see Section
  \ref{sec:vmin_regions}): the thin solid line where
  $v_{min}(E^{LUX}_{min})=v^{Na}_{min}(E^{DAMA}_{max})$; the thick
  solid line where
  $v_{min}(E^{LUX}_{min})=v^{Na}_{min}(E^{DAMA}_{min})$; the thin
  long--dashed line where $v^{Na}_{min}(E^{DAMA}_{max})=v_{esc}$; the
  thick long--dashed line where
  $v^{Na}_{min}(E^{DAMA}_{min})=v_{esc}$. The blue shaded strip
  represents points for which $\Delta_{ST}>1.64$, where $\Delta_{ST}$
  is the shape--test parameter defined in
  Eq.(\protect\ref{eq:shape_test}). The corresponding boundaries for
  XENON100 are less constraining and lie outside the boundaries of the
  figure. In all the shown $m_{DM}$--$\delta$ interval the $v_{min}$
  range corresponding to an explanation of the DAMA effect with WIMP
  scatterings off Iodine targets extends beyond the escape velocity.}
\label{fig:mchi_delta_na}
\end{figure}

\begin{figure}[h]
\begin{center}
\includegraphics[width=0.7\columnwidth,bb= 46 194 506 635]{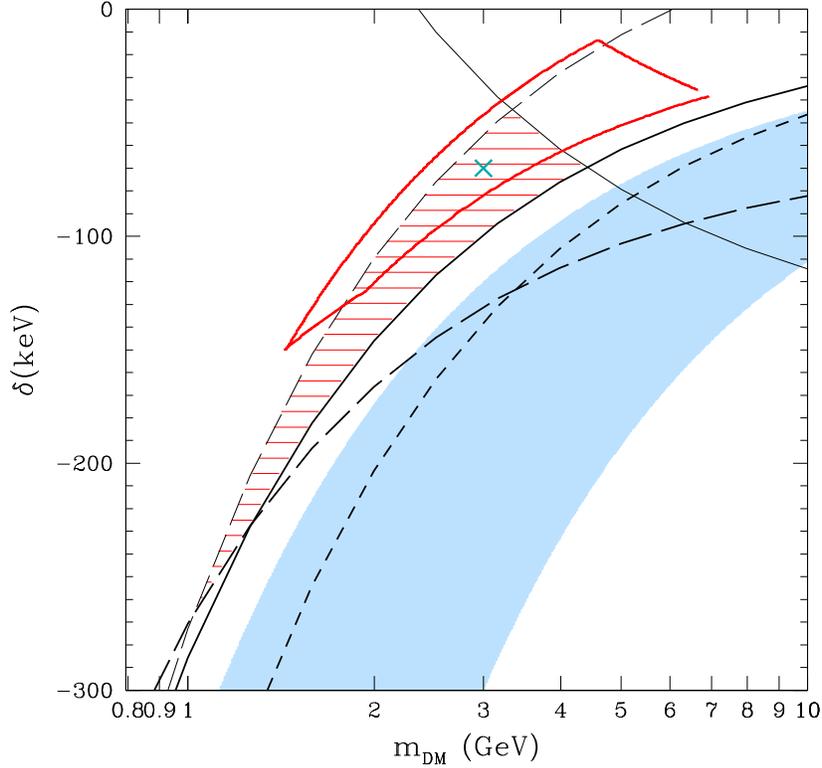}
\end{center}
\caption{Same as in Fig. \protect\ref{fig:mchi_delta_na} for the
  Silicon target in CDMS \protect\cite{cdms_si}. The horizontally
  (red) hatched area represents the IDM parameter space where the
  excess measured by CDMS-$Si$ corresponds to a $v_{min}<v_{esc}$
  range which is always below the corresponding one probed by LUX and
  XENON100. As explained in the text, in this case Xenon experiments
  can constrain the CDMS-$Si$ excess only when some assumptions are
  made on the galactic velocity distribution. The enclosed region is
  the result of the combination of four conditions: the thin solid
  line where $v_{min}(E^{LUX}_{min})$=$v_{min}(E^{CDMS-Si}_{max})$;
  the thick solid line where
  $v_{min}(E^{LUX}_{min})$=$v_{min}(E^{CDMS-Si}_{min})$; the thin
  long--dashed line where $v_{min}(E^{CDMS-Si}_{max})$=$v_{esc}$; the
  thick long--dashed line where
  $v_{min}(E^{CDMS-Si}_{min})$=$v_{esc}$. The corresponding boundaries
  for XENON100 are less constraining: in particular the thin
  short--dashed line represents the parameter space where
  $v_{min}(E^{XENON100}_{min})$=$v_{min}(E^{CDMS-Si}_{min})$. The blue
  shaded strip represents points excluded by the mirror test
  introduced in Section \protect\ref{sec:shape_test} where $p<0.05$,
  with $p$ defined in Eq.(\protect\ref{eq:p_value_poisson}). The
  closed solid (red) contour is the same compatibility region shown in
  Fig. \protect\ref{fig:mchi_delta_na}. The cross represents the
  benchmark point whose $v_{min}$--$\tilde{\eta}_{0,1}$ parameter
  space is discussed in Figs
  \protect\ref{fig:benchmark_mchi_3_delta_m70}.}
\label{fig:mchi_delta_si}
\end{figure}

\begin{figure}[h]
\begin{center}
\includegraphics[width=0.7\columnwidth,bb= 46 194 506 635]{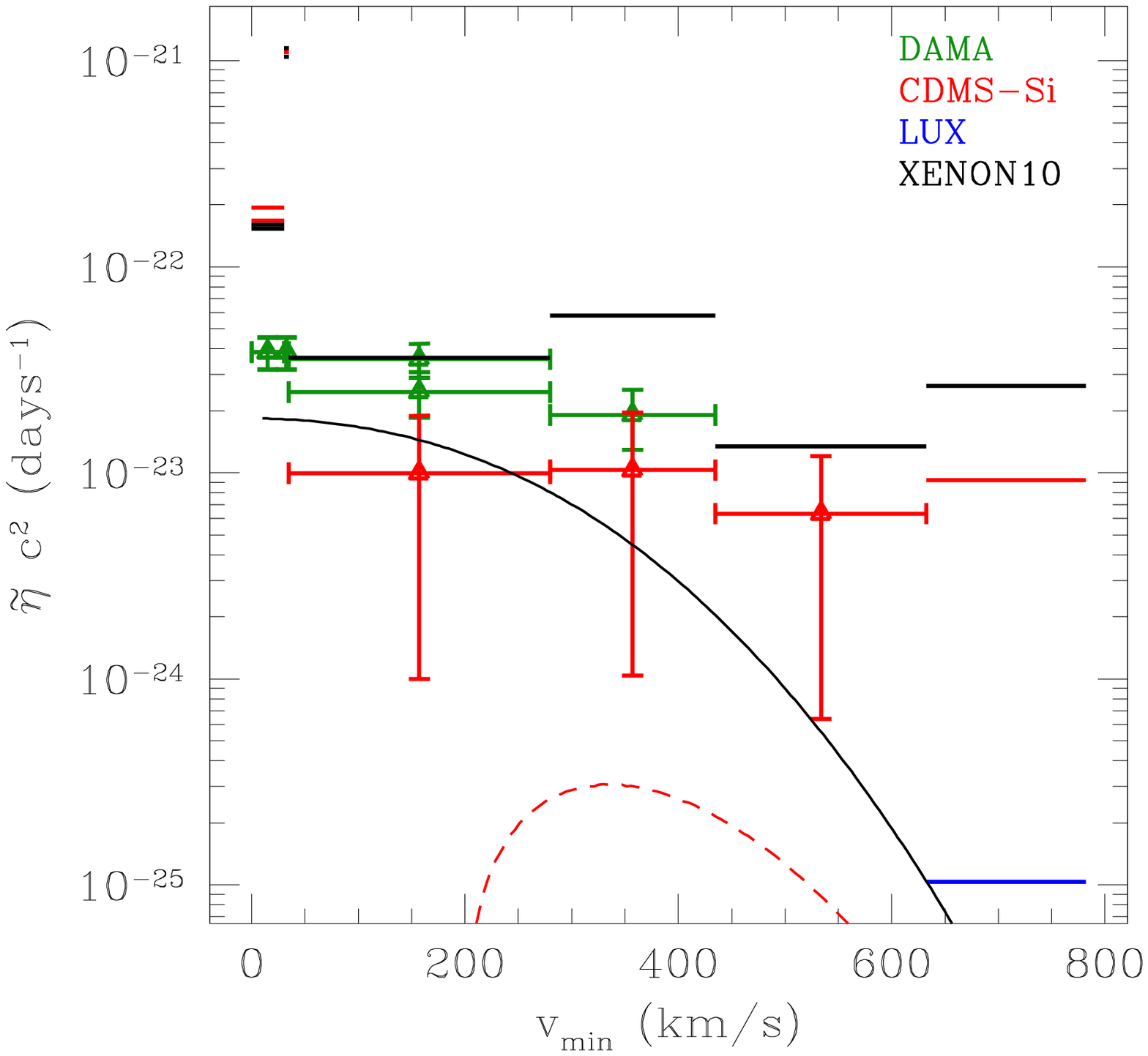}
\end{center}
\caption{Measurements and bounds for the functions $\tilde{\eta}_0$ and
  $\tilde{\eta}_1$ for the benchmark point $m_{DM}=3$ GeV,
  $\delta$=-70 keV, represented with a cross in
  Fig.\ref{fig:mchi_delta_si}, and assuming isospin violation
  $f_n/f_p$=-0.79. The solid (black) and dashed (red) lines
  represent the predictions for $\tilde{\eta}_0$ and $\tilde{\eta}_1$,
  respectively, for a Maxwellian velocity distribution, normalized to
  the most constraining upper bound.}
\label{fig:benchmark_mchi_3_delta_m70}
\end{figure}

Let us indicate with \verb!V_DAMA_NA!$\equiv
[v^{DAMA,Na}_{min},v^{DAMA,Na}_{max}]$ the $v_{min}$ range for the
DAMA signal assuming scattering on Sodium, and with
\verb!V_DAMA_I!$\equiv[v^{DAMA,I}_{min},v^{DAMA,I}_{max}]$ the
corresponding one for scattering on Iodine.
\verb!V_LUX!$\equiv[v^{LUX}_{min},v^{LUX}_{max}]$ represents the range
to which LUX is sensitive, while
\verb!V_XENON100!$\equiv[v^{XENON100}_{min},v^{XENON100}_{max}]$ is
the same for XENON100 (see appendix \ref{app:exp} for experimental details).

The result of our analysis is shown in Fig. \ref{fig:mchi_delta_na},
where we use $Q_{Na}$=0.3 for the Sodium quenching factor (see
Appendix \ref{app:exp}). In the whole $m_{DM}$--$\delta$ range we find
\verb!V_DAMA_I!$\cap$\verb!V_GAL!=0, so that the assumption of
dominance of scatterings off Sodium is consistent. Moreover in the
region between the two long--dashed lines \verb!V_DAMA_NA!  $\subset$
\verb!V_GAL!  , i.e. the required values of $v_{min}$ to explain the
signal with scatterings on Sodium are all below $v_{esc}$: in
particular, the thick long--dashed line corresponds to
$v_{min}(E_{min}^{Na})=v_{esc}$, while the thin long--dashed line to
$v_{min}(E_{max}^{Na})=v_{esc}$.  In the same figure the (light--blue)
shaded area is excluded by the shape test described in Section
\ref{sec:shape_test} (the shaded area extends beyond the region with
\verb!V_DAMA_NA! $\subset$ \verb!V_GAL!  because in that domain only a
fraction of the signal interval is automatically selected to do the
test, while the remaining part, or a fraction of it, corresponds to
$v_{min}$ values larger than $v_{esc}$).  Moreover, in the region
above the thick solid line (corresponding to
$v_{min}(E_{min}^{Na})=v_{min}(E_{min}^{LUX})$) and below the thin
solid line (corresponding to
$v_{min}(E_{max}^{Na})=v_{min}(E_{min}^{LUX})$) one has \verb!V_LUX!
$>$ \verb!V_DAMA_NA!. Notice that in both cases the boundaries are
determined by the light response of LUX at threshold, and that this
issue for liquid dual--phase scintillators is still controversial
\cite{collar_liquid}. On the other hand the same boundaries from
XENON100 are less constraining (due to the higher lower threshold) and
lie outside the plot. The overlapping of the two regions
\verb!V_DAMA_NA! $\subset$ \verb!V_GAL! and \verb!V_LUX! $>$
\verb!V_DAMA_NA! is not affected by the shape test and in
Fig. \ref{fig:mchi_delta_na} is given by the horizontally (red)
hatched area. In this region LUX can constrain the DAMA excess only if
our knowledge on the $\tilde{\eta}_0$ function allows us to
extrapolate it from the LUX range down to the DAMA-$Na$ range. If on
the other hand no assumptions are made about $\tilde{\eta}_0$ besides
(\ref{eq:eta_conditions}) the two results can be made compatible for a
wide range of $\tilde{\eta}_0$ functional forms. Notice that this
argument involves exclusively kinematics, and is valid no matter what
the dynamics of the process is. This means in particular that it would
hold also if the scaling law of the WIMP--nucleus cross section were
different from the one given in Eq.(\ref{eq:scaling_law}), and/or the
dynamics were modified by some other effect, such as, for instance,
the exchange of a light mediator or a magnetic--type coupling,
introducing a dependence of the differential rate on the recoil energy
and/or the incoming WIMP velocity different from the one given in
Eq.(\ref{eq:diff_rate_er}) \footnote{As discussed in
  \protect\cite{gondolo_eta_generalized} in the case of a generalized
  interaction a factorizable definition of the $\eta$ function
  different from Eq.(\protect\ref{eq:eta}) would still be possible.}.

The same check can be made between DAMA scatterings on Sodium and the
SuperCDMS experiment bound\cite{super_cdms} which uses a Germanium
target. Proceeding as before, indicating by
\verb!V_SUPERCDMS!$\equiv[v^{SuperCDMS}_{min},v^{SuperCDMS}_{max}]$
the corresponding $v_{min}$ interval, we have checked that
\verb!V_SUPERCDMS!$\le$\verb!V_DAMA! over all the $m_{DM}$--$\delta$
range of Eq. (\ref{eq:mchi_delta_ranges}), i.e. SuperCDMS is always
sensitive to $v_{min}$ values in the same range or smaller than those
which could explain the DAMA effect.  This means that, besides
experimental issues, the two measurements cannot be reconciled using
kinematics arguments only. However, in presence of some additional
dynamical mechanism suppressing WIMP scatterings on Germanium compared
to that on Sodium, DAMA and SuperCDMS can in principle be
reconciled. An example of such mechanism is the Isospin violation
mechanism\cite{isospin_violation}, where a specific choice of the
$f_n/f_p$ ratio in Equation (\ref{eq:scaling_law}) can suppress the
WIMP coupling to targets within a restricted range of Atomic numbers:
in particular, by choosing $f_n/f_p\simeq$
-0.79\footnote{Next--to--leading order corrections in the chiral
  expansion of the effective WIMP--quark interaction can modify the
  value of the $f_n/f_p$ ratio for which the cancellation takes
  place\cite{nlo_cirigliano}.} the SuperCDMS bound (as well as those
of all other experiments using Ge targets) would no longer be present.

In Fig.\ref{fig:mchi_delta_si} we repeat the same analysis by
comparing the excess of three events claimed by the CDM-$Si$
experiment\cite{cdms_si} to LUX, XENON100 and SuperCDMS. The result
turns out to be qualitatively similar to the previous case. By
indicating with
\verb!V_CDMS_SI!$\equiv[v^{CDMS-Si}_{min},v^{CDMS-Si}_{max}]$ the
$v_{min}$ interval that can explain the CDMS--$Si$ excess, in
Fig. \ref{fig:mchi_delta_si} the region between the two long--dashed
lines has \verb!V_CDMS_SI!$\subset$\verb!V_GAL! (the thick
long--dashed line corresponds to $v_{min}(E_{min}^{CDMS-Si})=v_{esc}$
while the thin long--dashed line to
$v_{min}(E_{max}^{CDMS-Si})=v_{esc}$). On the other hand in the region
above the thick solid line (corresponding to
$v_{min}(E_{min}^{CDMS-Si})=v_{min}(E_{min}^{LUX})$) and below the
thin solid line (corresponding to
$v_{min}(E_{max}^{CDMS-Si})=v_{min}(E_{min}^{LUX})$) one has
\verb!V_LUX!  $>$ \verb!V_CDMS_SI!. Similarly to what happens in
Fig.\ref{fig:mchi_delta_na} the same boundaries for XENON100 are less
constraining due the higher energy threshold: however in this case the
curve corresponding to
$v_{min}(E_{min}^{CDMS-Si})=v_{min}(E_{min}^{XENON100})$ lies within
the plot and is represented by the thick short--dashed line.  The
overlapping of the two regions \verb!V_CDMS_SI!$\subset$\verb!V_GAL!
and \verb!V_LUX!  $>$ \verb!V_CDMS_SI!  is given by the horizontally
(red) hatched area and is not affected by the (light--blue) shaded
area excluded by the mirror test introduced in Section
\ref{sec:shape_test} (notice that in the case of the three events in
CDMS-$Si$ the shape--test is obviously not as constraining as in the
DAMA case).

Similarly to what happened in Fig.\ref{fig:mchi_delta_na} also in the
case of Fig.\ref{fig:mchi_delta_si} one has \verb!V_SUPERCDMS! $\le$
\verb!V_CDMS_SI! over all the shown $m_{DM}$--$\delta$ range. Again, a
possibility to reconcile CDMS--$Si$ and SuperCDMS is to assume the
same suppression mechanism for WIMP scattering on Ge advocated in the
discussion of Fig.\ref{fig:mchi_delta_na}.  Actually, in
Fig. \ref{fig:mchi_delta_si} the closed solid (red) contour
represents the same compatibility region shown in
Fig.\ref{fig:mchi_delta_na} for DAMA--$Na$.  Since the two regions
overlap, one may wonder whether compatibility among the two excesses
(DAMA--$Na$ and CDMS-$Si$) can actually be achieved in compliance with
constraints from other null results. For this reason in
Fig.\ref{fig:benchmark_mchi_3_delta_m70} we plot the measurements and
bounds on the functions $\tilde{\eta}_0$ and $\tilde{\eta}_1$ for the
benchmark choice $m_{DM}=3$ GeV, $\delta$=-70 GeV (corresponding to
the cross plotted in Fig.\ref{fig:mchi_delta_si}) and for
$f_n/f_p\simeq$ -0.79 to comply with the SuperCDMS bound. As shown in
the Figure, indeed, the DAMA-$Na$ and CDMS-$Si$ can be separately brought
into agreement with other experimental constraints for the same choice
of $m_{DM}$ and $\delta$. In the same figure the solid (black) line
represents the $\tilde{\eta}_0$ function in the case of a Maxwellian
velocity distribution normalized to the LUX upper bound, and the
dashed (red) curve the corresponding prediction for $\tilde{\eta}_1$.
From this plot it is clear that for an Isothermal Sphere model even by
assuming isospin violation it is not possible to separately reconcile
DAMA and CDMS-$Si$ with both LUX and SuperCDMS, but this can in
principle achieved if no assumptions are made on the velocity
distribution.  However, even in this case DAMA and CDMS--$Si$ appear in
tension with each other, since the $\tilde{\eta}_1$ ranges explaining
DAMA-$Na$ are systematically larger than the $\tilde{\eta}_0$ ranges
required to explain the CDMS-$Si$ excess (in disagreement to the last of
conditions (\ref{eq:eta_conditions})).

Notice that in some of the $v_{min}$ intervals of
Fig. \ref{fig:benchmark_mchi_3_delta_m70} DAMA provides two
independent determinations of $\bar{\tilde{\eta}}_1$, a consequence of
the double mapping of Eq.(\ref{eq:vmin}), and that these
determinations appear to be in mutual agreement (as confirmed by the
fact that for this specific choice of $m_{DM}$ and $\delta$ we find
the shape test parameter value $\Delta_{ST}\simeq 0.95$).

\subsection{Iodine scattering in DAMA and the CRESST excess}
\label{sec:dama_i}

\begin{figure}[h]
\begin{center}
\includegraphics[width=0.7\columnwidth,bb= 46 194 506 635]{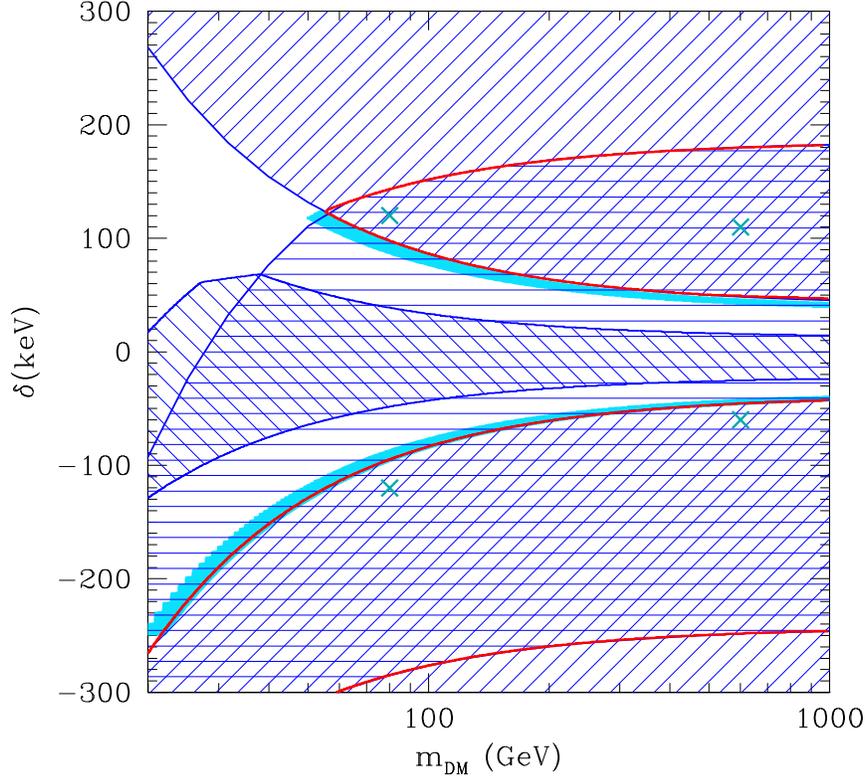}
\end{center}
\caption{Same as in Fig. \protect\ref{fig:mchi_delta_na} for the $I$
  target in DAMA \protect\cite{dama}. The region enclosed by the thick
  solid (red) line represents the IDM parameter space where the excess
  measured by DAMA corresponds to a $v_{min}<v_{esc}$ range which is
  always below the corresponding ones probed by LUX\protect\cite{lux}
  and SuperCDMS\protect\cite{super_cdms}. The enclosed region is the
  result of the combination of three conditions: in the region with
  horizontal hatches the whole $v_{min}$ range corresponding to the
  DAMA signal is below $v_{esc}$; +45$^{\circ}$ oblique hatches
  correspond to the domain where the $v_{min}$ range probed by LUX is
  at higher values compared to the range explaining DAMA; in the
  region with -45$^{\circ}$ oblique hatches the $v_{min}$ range probed
  by SuperCDMS overlaps or is at lower values compared to the range
  explaining DAMA. The blue shaded strip represents points where
  $\Delta_{ST}>1.64$, where $\Delta_{ST}$ is the shape-test parameter
  defined in Eq. (\protect\ref{eq:shape_test}). In all the shown
  $m_{DM}$--$\delta$ interval the $v_{min}$ range corresponding to an
  explanation of the DAMA effect with WIMP scatterings off Sodium
  targets extends beyond the escape velocity. The four crosses are the
  benchmark points whose $v_{min}$--$\tilde{\eta}_{0,1}$ parameter
  space is discussed in Figs. \protect\ref{fig:benchmarks_i}(a-d).}
\label{fig:mchi_delta_i}
\end{figure}

\begin{figure}[h]
\begin{center}
\includegraphics[width=0.49\columnwidth,bb= 20 178 504 630]{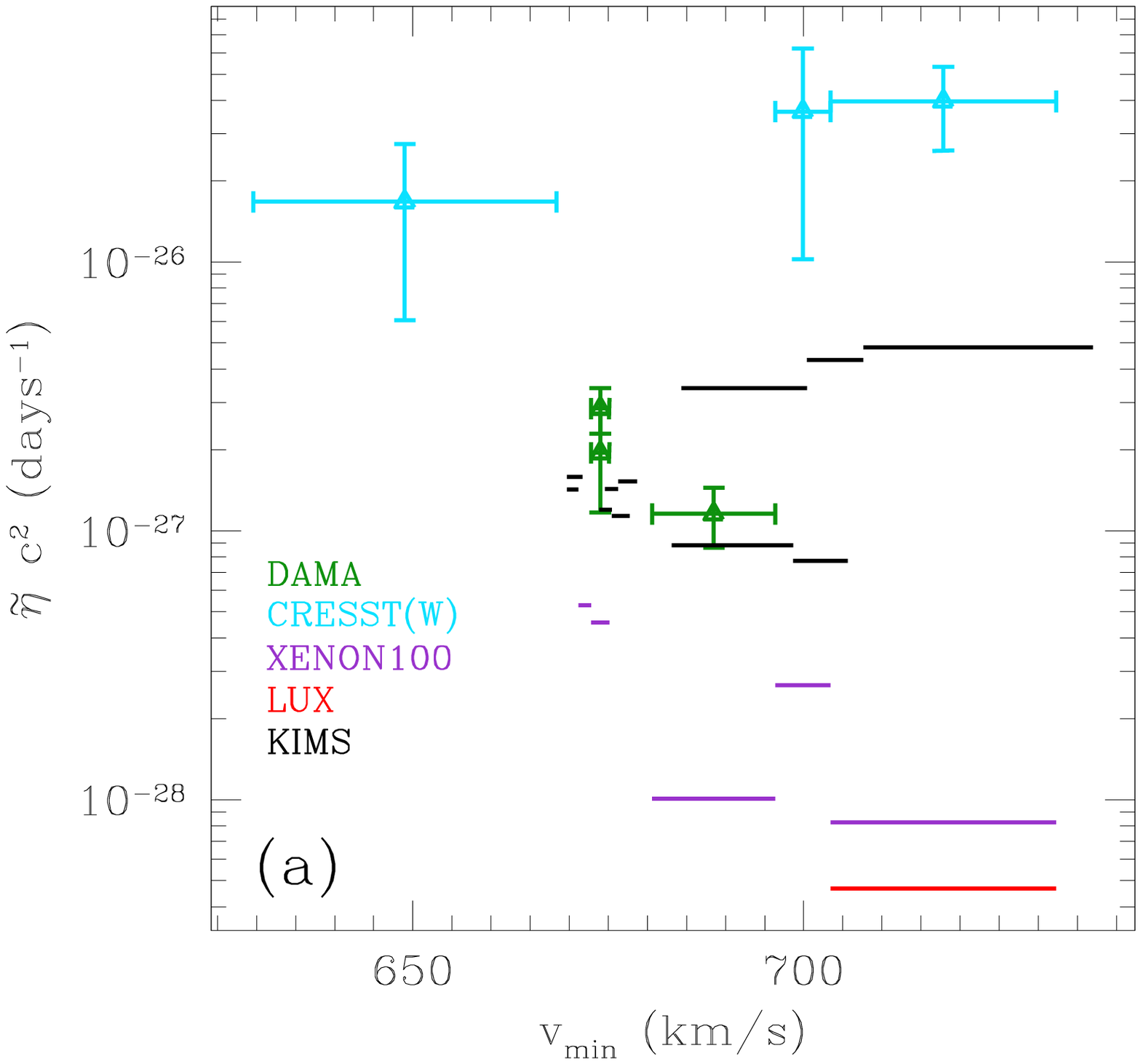}
\includegraphics[width=0.49\columnwidth,bb= 20 178 504 630]{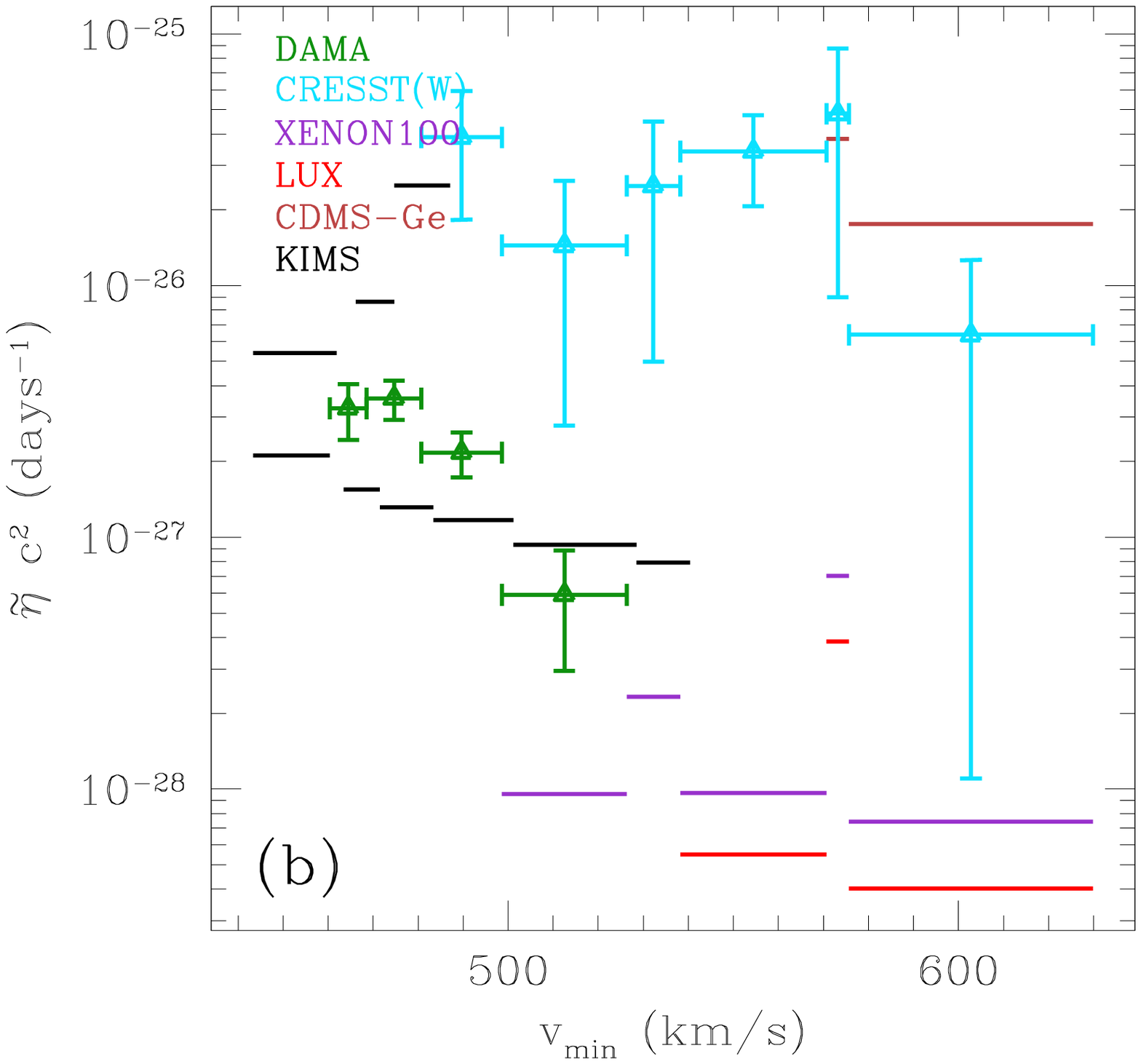}
\includegraphics[width=0.49\columnwidth,bb= 20 178 504 630]{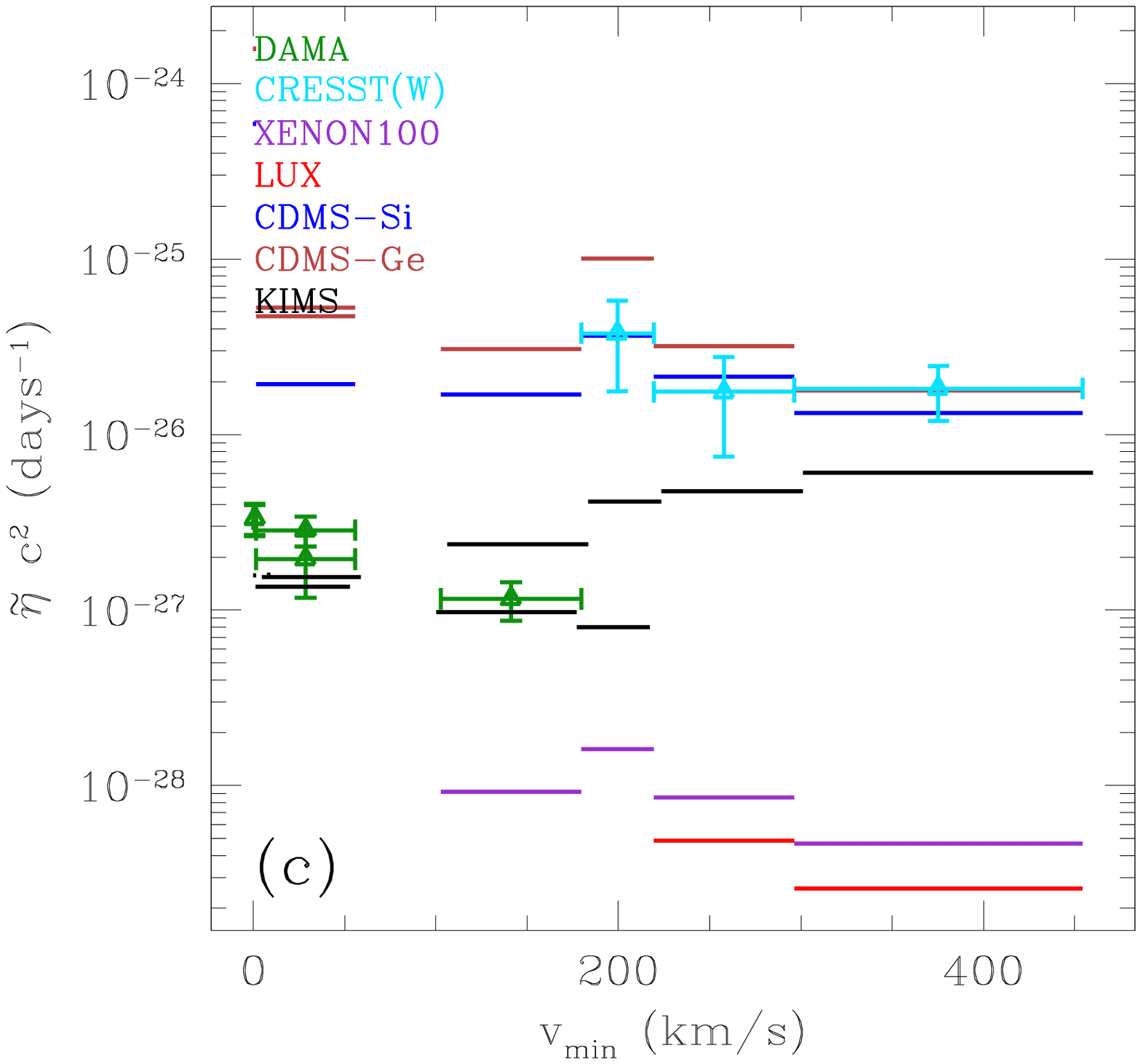}
\includegraphics[width=0.49\columnwidth,bb= 20 178 504 630]{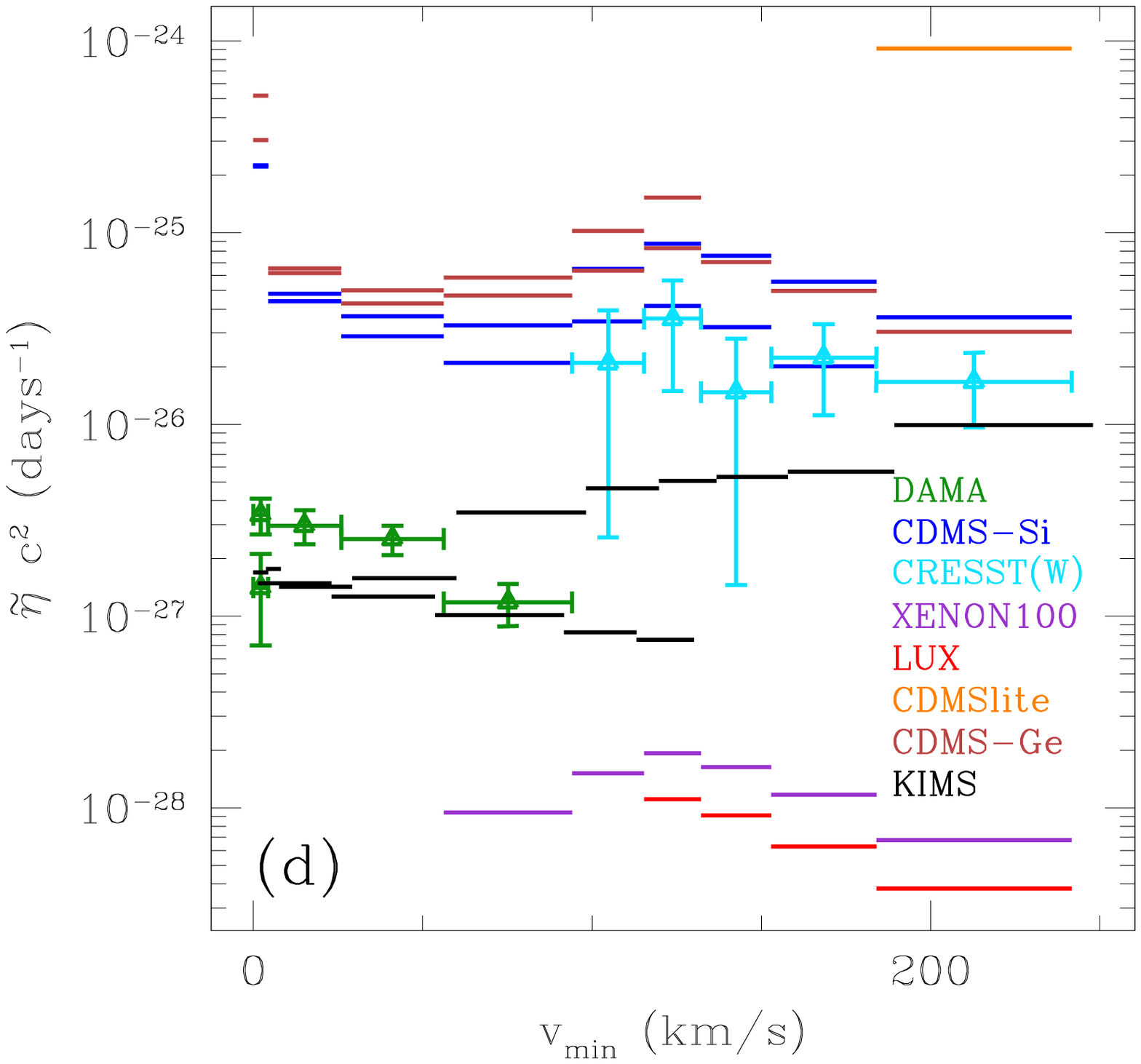}
\end{center}
\caption{Measurements and bounds for the functions $\tilde{\eta}_0$
  and $\tilde{\eta}_1$ for the four benchmark points represented with
  crosses in Fig.\ref{fig:mchi_delta_i} and for $f_n/f_p$=1. {\bf (a)}
  $m_{DM}=$80 GeV, $\delta$=120 keV; {\bf (b)} $m_{DM}=$600 GeV,
  $\delta$=110 keV; {\bf (c)} $m_{DM}=$80 GeV, $\delta$=-120 keV; {\bf
    (d)} $m_{DM}=$600 GeV, $\delta$=-60 keV. In all these benchmarks
  the tension between DAMA and other experiments is maximally
  alleviated, since it is reduced to that with the minimal number of
  other experiments (KIMS and XENON100). }
\label{fig:benchmarks_i}
\end{figure}

\begin{figure}[h]
\begin{center}
\includegraphics[width=0.7\columnwidth,bb= 46 194 506 635]{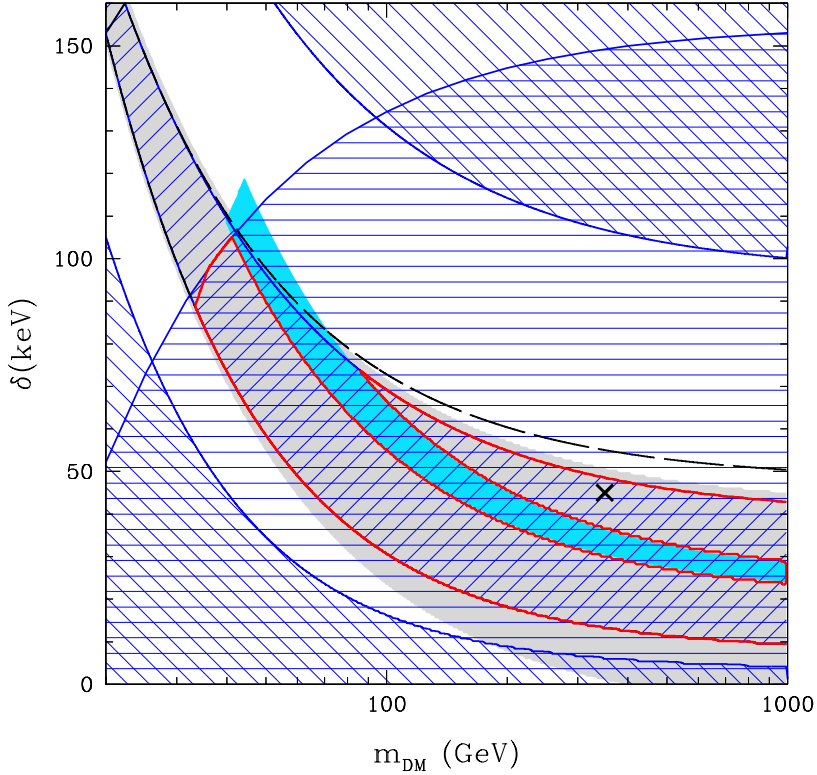}
\end{center}
\caption{Same as in Fig. \protect\ref{fig:mchi_delta_na} for the
  Tungsten target in CRESST \protect\cite{cresst}. The region enclosed
  by the thick solid (red) line represents the IDM parameter space
  where the excess measured by CRESST corresponds to a
  $v_{min}<v_{esc}$ range which is always below the corresponding ones
  probed by XENON100\protect\cite{xenon100},LUX\protect\cite{lux},
  SuperCDMS\protect\cite{super_cdms} and KIMS\protect\cite{kims}, and
  is not excluded by the mirror test introduced in Section
  \protect\ref{sec:shape_test}. The enclosed region is the result of
  the combination of four conditions: in the region with horizontal
  hatches the whole $v_{min}$ range corresponding to the CRESST signal
  is below $v_{esc}$; +45$^{\circ}$ oblique hatches correspond to the
  domain where the $v_{min}$ range probed by XENON100 is at higher
  values compared to the range explaining CRESST (the bound from LUX
  is slightly less constraining and the upper part of the
  corresponding closed region is represented by the dashed (black)
  line); in the region with -45$^{\circ}$ oblique hatches the
  $v_{min}$ range probed by SuperCDMS overlaps or is at lower values
  compared to the range explaining CRESST; the light (gray) shaded
  area corresponds to the domain where the $v_{min}$ range probed by
  KIMS is at higher values compared to the range explaining
  CRESST. The blue shaded strip represents points where where
  $p<0.05$, with $p$ defined in
  Eq.(\protect\ref{eq:p_value_poisson}). In this figure $v_{min}$
  ranges corresponding to WIMP scatterings off $Ca$ or $O$ targets are
  not always beyond $v_{esc}$ (see discussion of
  Fig. \ref{fig:mchi_delta_i_w}) so dominance of WIMP--$W$
  scatterings must be established dynamically. This is achieved, for
  instance, in the isospin--conserving case $f_n/f_p$=1, when the WIMP
  coupling to $W$ nuclei is much larger than that on $Ca$ and $O$ (see
  Fig. \ref{fig:scaling_laws}(a)). The cross represents the benchmark
  point whose $v_{min}$--$\tilde{\eta}_{0,1}$ parameter space is
  discussed in Fig.  \protect\ref{fig:benchmark_mchi_350_delta_45}.}
\label{fig:mchi_delta_w}
\end{figure}

\begin{figure}[h]
\begin{center}
\includegraphics[width=0.7\columnwidth,bb= 46 194 506 635]{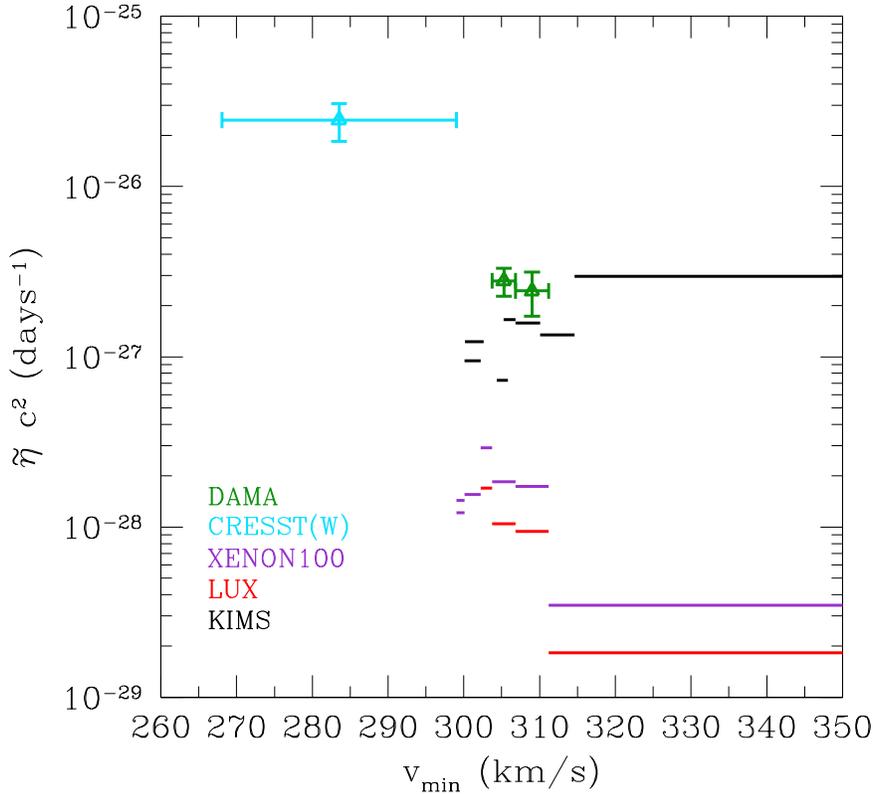}
\end{center}
\caption{Measurements and bounds for the functions $\tilde{\eta}_0$ and
  $\tilde{\eta}_1$ for the benchmark point $m_{DM}=350$ GeV,
  $\delta$=45 keV, represented with a cross in
  Fig.\ref{fig:mchi_delta_w}(a), and assuming an isospin conserving
  coupling, $f_n/f_p$=1.}
\label{fig:benchmark_mchi_350_delta_45}
\end{figure}

\begin{figure}[h]
\begin{center}
\includegraphics[width=0.7\columnwidth,bb= 46 194 506 635]{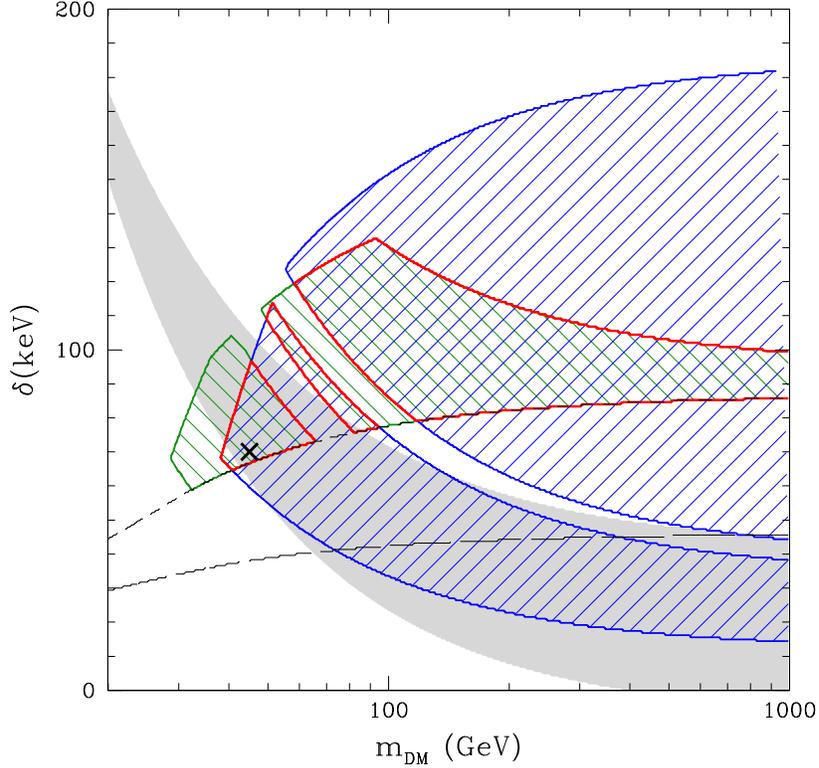}
\end{center}
\caption{Superposition of allowed regions for WIMP scattering off
  Iodine in DAMA (+45$^{\circ}$ blue--hatched area) and off Tungsten
  in CRESST (-45$^{\circ}$ green--hatched area), when the Xenon bounds
  (LUX and XENON100) are not included.  If some mechanism is advocated
  to suppress WIMP couplings on Xenon targets the region enclosed by
  the thick (red) line represents the parameter space where both the
  DAMA and CRESST effects correspond to $v_{min}<v_{esc}$ ranges which
  are always below the corresponding one probed by
  SuperCDMS\protect\cite{super_cdms}. Specifically, if isospin
  violation, $f_n/f_p\simeq$-0.69, is assumed to suppress WIMP
  couplings on $Xe$ targets, the coupling of WIMPs to $Ca$ is enhanced
  compared to that on $W$ (see Fig. \ref{fig:scaling_laws})(b) so in
  order to assume dominance of WIMP--$W$ scatterings the region above
  the curve with short-dashes must be considered, where the $v_{min}$
  range corresponding to WIMP--$Ca$ scatterings is beyond $v_{esc}$
  (the region above the line with long-dashes corresponds to the same
  condition for WIMP--$O$ scatterings). Finally the light (gray)
  shaded area corresponds to the domain where the $v_{min}$ range
  probed by KIMS is at higher values compared to the range explaining
  CRESST.  The cross represents the benchmark point whose
  $v_{min}$--$\tilde{\eta}_{0,1}$ parameter space is discussed in Fig.
  \protect\ref{fig:benchmark_mchi_45_delta_70}.}
\label{fig:mchi_delta_i_w}
\end{figure}

\begin{figure}[h]
\begin{center}
\includegraphics[scale=0.5,bb= 57 45 340 743,angle=-90,clip=true]{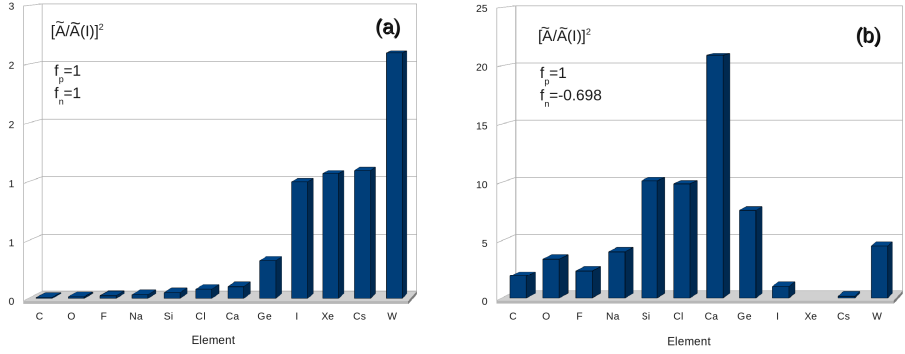}
\end{center}
\caption{Scaling law defined in Eq.(\ref{eq:scaling_law}) (normalized
  to that for WIMP--$I$ scattering) for the nuclear targets considered
  in the present analysis. {\bf (a)} Isospin conserving case
  ($f_n/f_p$=1). {\bf (b)} Isospin violating case $f_n/f_p$=-0.698,
  corresponding to the maximal suppression of the WIMP coupling to
  $Xe$ targets.}
\label{fig:scaling_laws}
\end{figure}

\begin{figure}[h]
\begin{center}
\includegraphics[width=0.7\columnwidth,bb= 46 194 506 635]{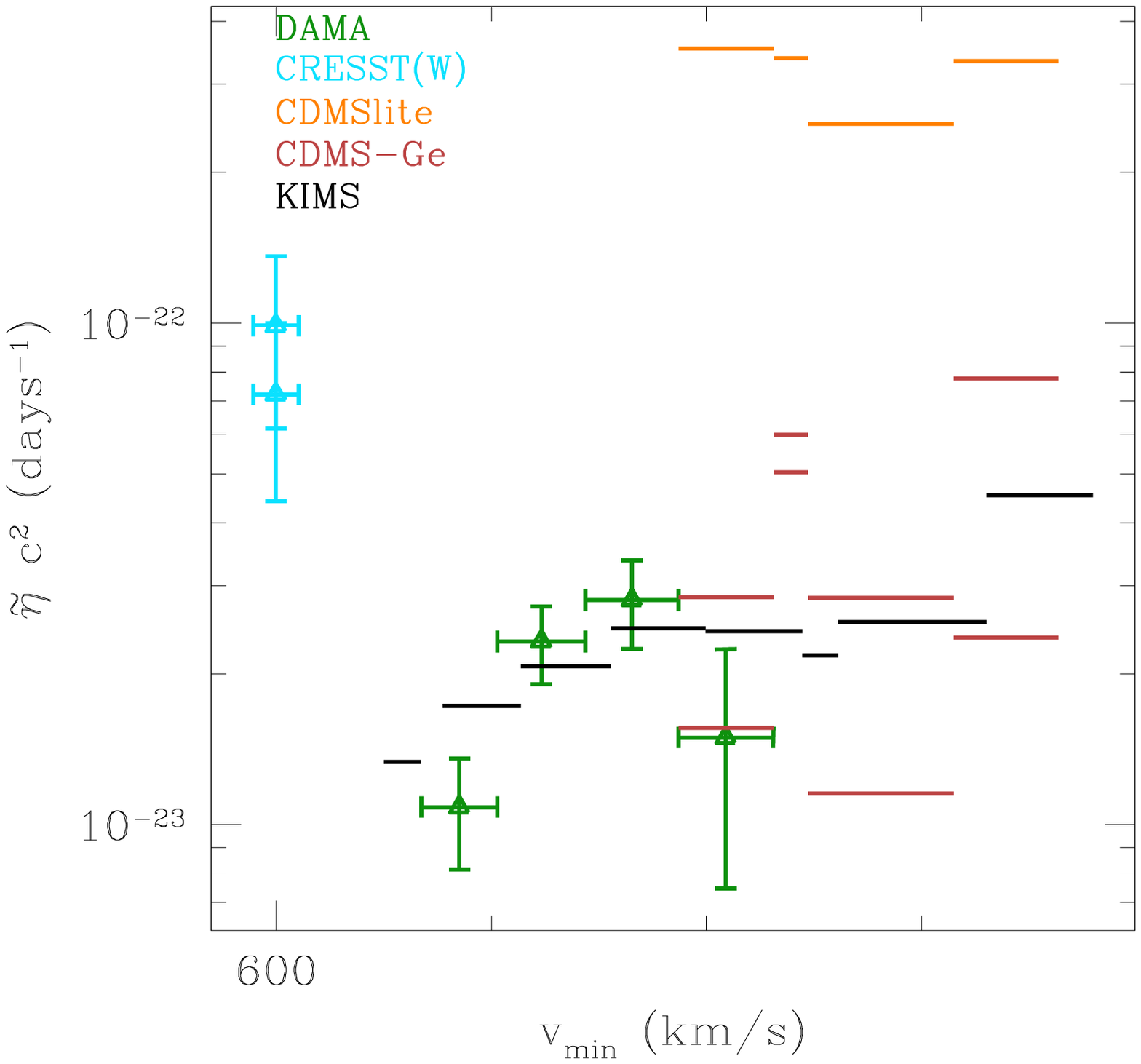}
\end{center}
\caption{Measurements and bounds for the functions $\tilde{\eta}_0$ and
  $\tilde{\eta}_1$ for the benchmark point $m_{DM}=45$ GeV,
  $\delta$=70 keV, represented with a cross in
  Fig.\ref{fig:mchi_delta_i_w}, and assuming an isospin violating
  coupling, $f_n/f_p$=-0.69}
\label{fig:benchmark_mchi_45_delta_70}
\end{figure}

In this section we will assume Iodine scattering in DAMA and Tungsten
scattering in CRESST. As already pointed out previously and shown in
Fig. \ref{fig:benchmark_mchi_100_delta_0}, a tension arises in this
case between DAMA and KIMS and since both experiments use the same
Iodine target nuclei this discrepancy cannot be solved by changing the
particle--physics model of the WIMP interaction. On the other hand, as
we will see, the tension with at least some of the other experiments
can in this case be alleviated by a combination of particle physics
assumptions and by allowing a generalized WIMP velocity distribution
complying with the minimal set of assumptions summarized in
Eq. (\ref{eq:eta_general}).

The result of a scanning of the $m_{DM}$--$\delta$ parameter space
assuming Iodine scattering in DAMA is shown in
Fig. \ref{fig:mchi_delta_i}. In all the shown range
\verb!V_DAMA_NA!$\cap$\verb!V_GAL!=0 while in the horizontally hatched
domain \verb!V_DAMA_I! $\subset$ \verb!V_GAL!.  Moreover, the
(light--blue) shaded areas are excluded by the shape test.  In this
case DAMA regions compatible to both LUX and SuperCDMS, can be
found. In particular, in the +45 $^{\circ}$ hatched region
\verb!V_LUX! $>$ \verb!V_DAMA_I!, while in the -45$^{\circ}$ hatched
area \verb!V_SUPERCDMS! $<$ \verb!V_DAMA_I!, i.e., in this case the
marked region corresponds to the excluded one. However, in the same
figure no regions of compatibility between DAMA and XENON100 can be
found . This result is in agreement to the analysis of
Ref.\cite{halo_independent_inelastic}. One comment is in order
here. The boundary of the +45$^{\circ}$ hatched region in
Fig. \ref{fig:mchi_delta_i} is determined by a combination of the two
conditions: $v_{min}(E^{LUX}_{max})=v_{min}(E^{DAMA}_{min})$ and
$v_{min}(E^{LUX}_{max})=v_{min}(E^{DAMA}_{max})$, i.e. the boundary of
that region is determined by the upper edge of the LUX analyzed energy
interval ($S1$=30 PE, see Appendix \ref{app:exp}).  In this Section we
adopt the value $Q_I$=0.07 for the Iodine quenching factor (see
Appendix \ref{app:exp}). This value, which is within the large
systematic uncertainties on $Q_I$, reduces the tension between DAMA
and LUX by increasing $E^{DAMA}_{min}$=2 keV$/Q_I$, reducing
$v_{min}(E^{DAMA}_{min})$ and making the condition
$v_{min}(E^{LUX}_{max})>v_{min}(E^{DAMA}_{min})$ easier to achieve.
Notice also that the upper end of the analyzed spectrum in liquid
Xenon detectors is chosen to avoid the background from cosmogenic
$^{127}Xe$\cite{lux_127xe}, so is common also to XENON100.  However,
mainly due to the larger light yield in LUX ($L_y$=8.8 PE) compared to
that in XENON100 ($L_y$=2.28 PE) when converted in KeVnr the upper
bound of the XENON100 Region of Interest is significantly larger than
that for LUX (adopting the experimental inputs summarized in Appendix
\ref{app:exp} one gets $E_R^{max}$=43.04 keVnr for XENON100
vs. $E_R^{max}$=24.9 keVnr for LUX). For this reason in this case
XENON100 is more constraining than LUX. Notice that since the higher
part of the analyzed spectrum in XENON100 is not fraught by the many
systematic uncertainties which characterize the region close to
threshold\cite{collar_liquid} one should expect the ensuing
constraints to be robust. However, in the spirit of minimizing the
tension between DAMA and other experiments by reducing it to that with
the minimal number of other experiments (in this case KIMS and
XENON100), in Fig.\ref{fig:mchi_delta_i} we single out the regions
where DAMA is compatible with LUX and SuperCDMS with a thick (red)
boundary. Within those boundaries we select four benchmark points
marked by crosses in Figure \ref{fig:mchi_delta_i} and we adopt them
to analyze the experimental data, getting estimations of the
$\tilde{\eta}_0$ and $\tilde{\eta}_1$ functions in Figure
\ref{fig:benchmarks_i}(a-d) (assuming $f_n/f_p$=1). In all the plots
of Fig. \ref{fig:benchmarks_i} we also show the ranges of the
$\tilde{\eta}_0$ function which can explain the CRESST effect, where
scatterings on Tungsten are assumed to be dominant. Notice that while,
as expected in all the plots of Fig. \ref{fig:benchmarks_i}(a-d) the
DAMA points are compatible with the other constraints (except KIMS and
XENON100), in these particular benchmarks points CRESST is
incompatible with LUX and XENON100.

The corresponding analysis of the IDM parameter space where the CRESST
excess is assumed to be explained with WIMP scatterings off Tungsten
targets is shown in Fig.\ref{fig:mchi_delta_w}. Let us indicate with
\verb!V_CRESST_W!=$[v_{min}^{CRESST,W},v_{max}^{CRESST,W}]$ the
$v_{min}$ range where WIMP--$W$ scatterings correspond to the signal
region in CRESST. In Fig.\ref{fig:mchi_delta_w} the region with
horizontal hatches contains configurations for which \verb!V_CRESST_W!
$\subset$ \verb!V_GAL!. Tungsten is by large the heaviest among the
targets in $Ca W O_4$, so if, for instance, $f_n/f_p$=1, the scaling
law given by Eq.(\ref{eq:scaling_law}) implies that in this region we
can safely assume that $W$ scatterings dominate over other targets
(see Fig.\ref{fig:scaling_laws}(a)). Notice that the four benchmark
points selected in Fig.\ref{fig:mchi_delta_i} are all contained in
this domain, so that in all the plots of Fig.\ref{fig:benchmarks_i}
the assumption of $W$ domination is consistent. Moreover, in the
+45$^{\circ}$ hatched region of Fig.\ref{fig:mchi_delta_w}
\verb!V_XENON100!  $>$ \verb!V_CRESST_W! (the bound from LUX is
slightly less constraining and the upper part of the corresponding
closed region where \verb!V_LUX!  $>$ \verb!V_CRESST_W!  is
represented by the dashed (black) line), while in the -45$^{\circ}$
hatched area \verb!V_SUPERCDMS! $<$ \verb!V_CRESST_W!, i.e., also in
this case the marked region corresponds to the excluded one, and
XENON100 turns out to be more constraining than SuperCDMS.  Moreover,
indicating with \verb!V_KIMS!=$[v_{min}^{KIMS},v_{max}^{KIMS}]$ the
overall $v_{min}$ to which the KIMS experiment is sensitive, in the
same Figure the light (gray) shaded area represents the parameter
space where \verb!V_KIMS!$>$\verb!V_CRESST_W!, i.e. in that region the
CRESST excess is not constrained by KIMS.  Finally, the (light--blue)
shaded area is excluded by the mirror test, which now carves away a
part of the allowed region. In Fig.\ref{fig:mchi_delta_w} the two
remaining domains compatible to all the requirements and constraints
are marked by a thick (red) boundary. We then select within that
domain a benchmark point (indicated by a cross) whose corresponding
analysis in the $v_{min}$--$\tilde{\eta}_{0,1}$ space is shown in
Fig.\ref{fig:benchmark_mchi_350_delta_45}: indeed inspection of that
figure shows that in this case CRESST complies with all other
constraints.

In order to see if DAMA-$I$ and CRESST-$W$ can be explained by the
same IDM parameters and be compatible to the other experimental
constraints, in Fig.\ref{fig:mchi_delta_i_w} we superimpose the two
regions allowed by DAMA-$I$ and CRESST-$W$ when the LUX and XENON100
bounds are not applied. Again, in this case one needs to assume some
mechanism to suppress the WIMP coupling to $Xe$ targets, such as
$f_n/f_p\simeq$-0.69. However, as shown in the right panel of
Fig.\ref{fig:scaling_laws}, the particular choice $f_n/f_p\simeq$-0.69
not only suppresses the WIMP coupling to Xenon targets, but it also
enhances the same coupling to Calcium nuclei, so that in principle now
the condition \verb!V_CRESST_W!$\subset$\verb!V_GAL! no longer ensures
dominance of scatterings off $W$ nuclei in $CaWO_4$. However,
indicating with \verb!V_CRESST_CA! the $v_{min}$ interval for
scatterings off Ca nuclei, the region above the short--dashed line of
Fig.\ref{fig:mchi_delta_i_w} corresponds
\verb!V_CRESST_CA!$\cap$\verb!V_GAL!=0 (the region where the same
happens for scatterings off Oxygen targets is above the long--dashed
curve) so that in that domain $W$ dominance can be consistently
assumed. The DAMA--CRESST regions complying to all these bounds and
requirements are marked in Fig. \ref{fig:mchi_delta_i_w} by the thick
(red) boundaries. In particular one of the two regions overlaps with
the light (gray) shaded band where also the requirement
\verb!V_KIMS!$>$\verb!V_CRESST_W! is satisfied. The cross in the
latter domain represents a benchmark that we analyze in the
$v_{min}$--$\tilde{\eta}_{0,1}$ space in
Fig. \ref{fig:benchmark_mchi_45_delta_70}: inspection of that figure
shows that in this case compatibility between DAMA and CRESST is
achieved in compliance with SuperCDMS.  Moreover, although the tension
between DAMA and KIMS persists as expected, now the CRESST excess and
the KIMS bound are mutually compatible.

\subsection{Sodium scattering in DAMA at large WIMP masses}
\label{sec:na_large_mchi}

\begin{figure}[h]
\begin{center}
\includegraphics[width=0.7\columnwidth,bb= 46 194 506 635]{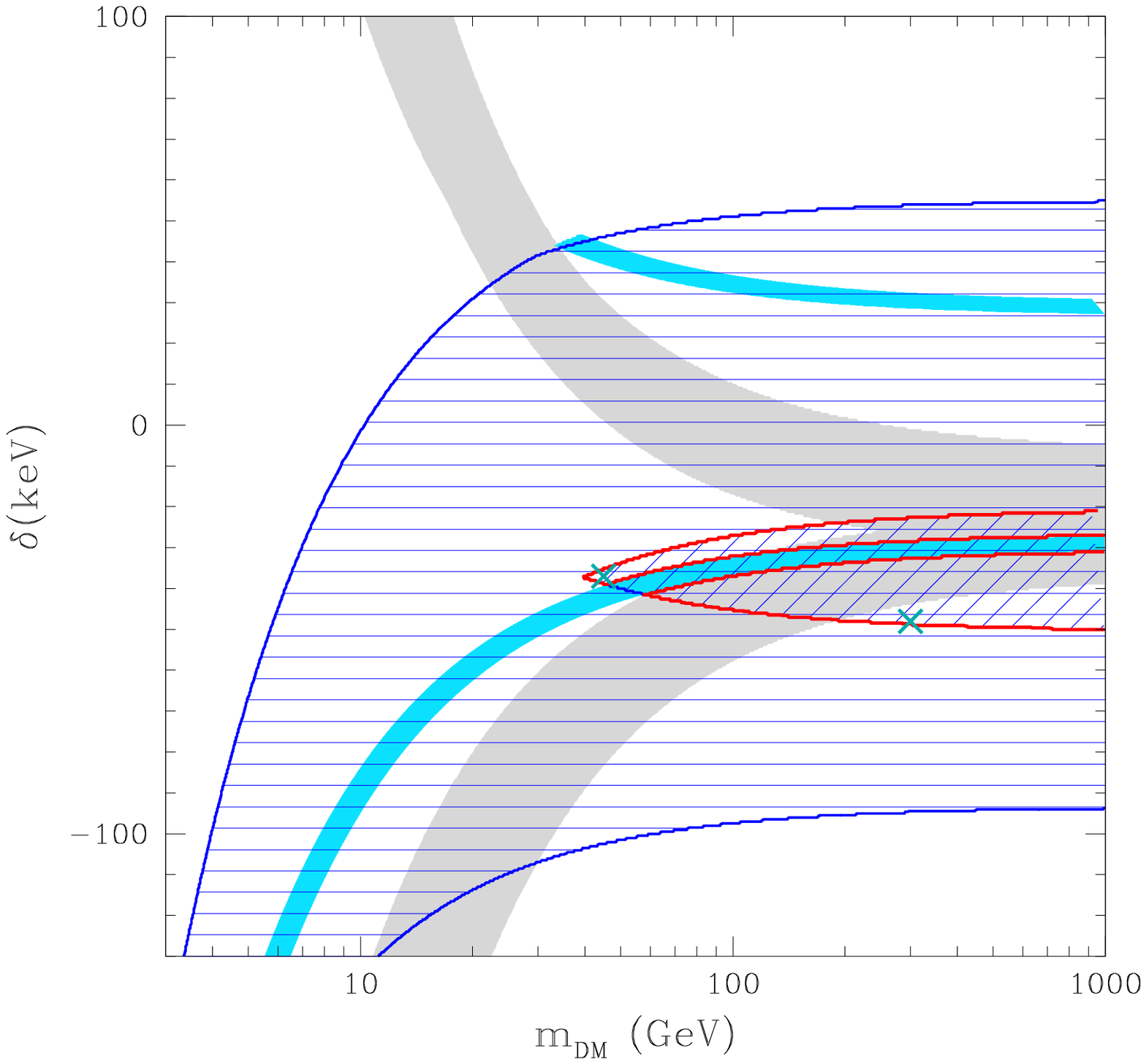}
\end{center}
\caption{Same as in Fig. \protect\ref{fig:mchi_delta_na} for the
  Sodium target in DAMA \protect\cite{cdms_si}, but assuming the $Na$
  quenching factor from Ref.\cite{quenching_collar} (see Appendix
  \ref{app:exp}). The two regions enclosed by the thick solid (red)
  line represent the IDM parameter space given by the combination of
  two requirements: (i) that the excess measured by DAMA corresponds
  to a $v_{min}$ range which is always below the corresponding one
  probed by SuperCDMS (the corresponding domain is represented by
  +45$^{\circ}$ hatches); (ii) the region is not excluded by the shape
  test introduced in Section \ref{sec:shape_test} (the shaded
  (light--blue) bands correspond to $\Delta_{ST}>$1.64). The area
  represented by horizontal hatches shows the parameter space where
  the $v_{min}<v_{esc}$ range explaining DAMA is all below
  $v_{esc}$. In this figure XENON detectors bounds are not included.
  The light (gray) shaded region represents the parameter space where
  the $v_{min}$ ranges mapped by scatterings off $Na$ and $I$ do not
  overlap: in this case dominance of scatterings off any of the two
  targets in $NaI$ is possible (see discussion in Section
  \ref{sec:na_large_mchi}). The two crosses are the benchmark points
  whose $v_{min}$--$\tilde{\eta}_{0,1}$ parameter space is discussed
  in Figs. \protect\ref{fig:benchmarks_na_quenching_collar}(a-b).}
\label{fig:mchi_delta_na_quenching_collar}
\end{figure}

\begin{figure}[h]
\begin{center}
\includegraphics[width=0.49\columnwidth,bb= 46 194 506 635]{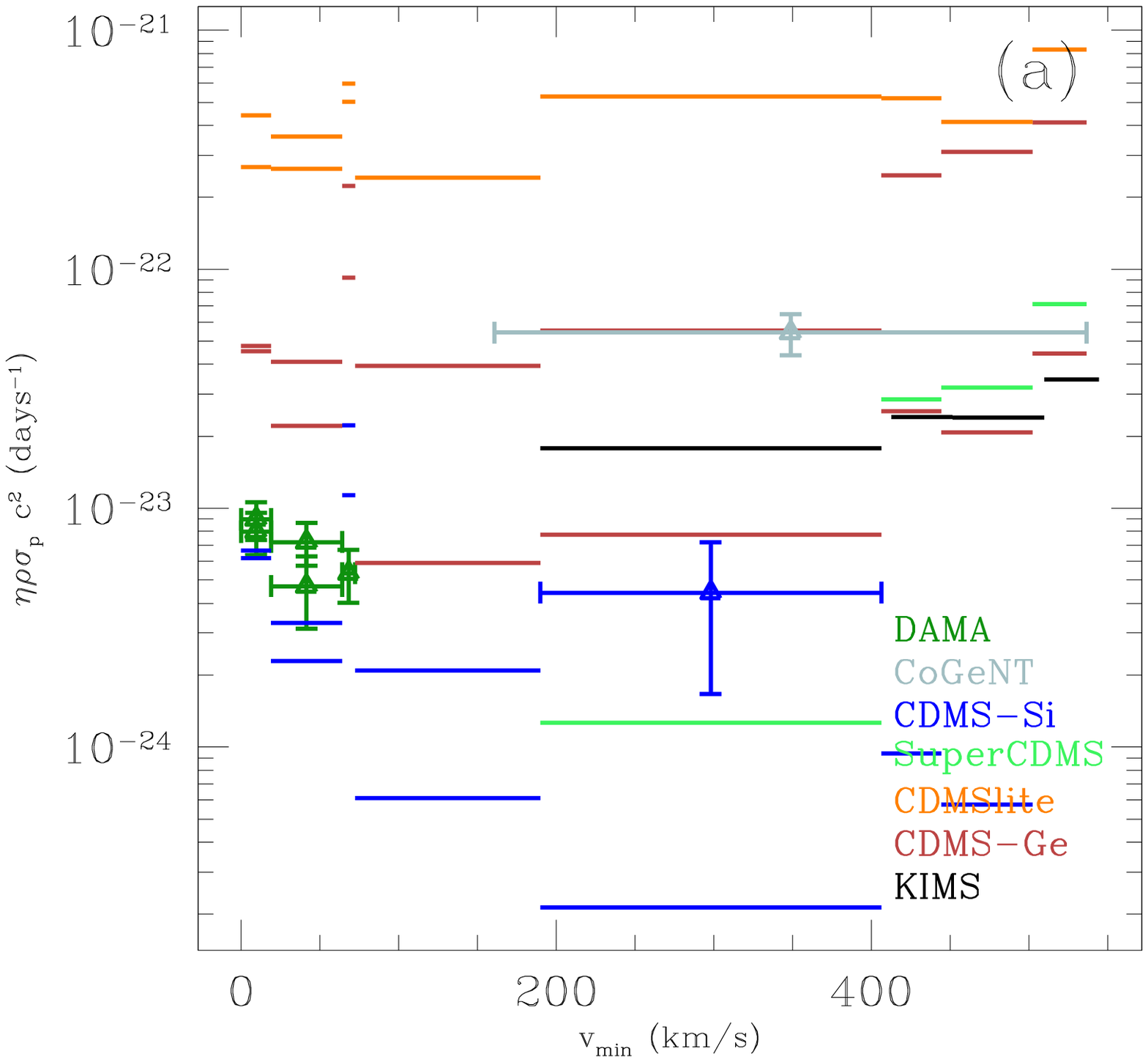}
\includegraphics[width=0.49\columnwidth,bb= 46 194 506 635]{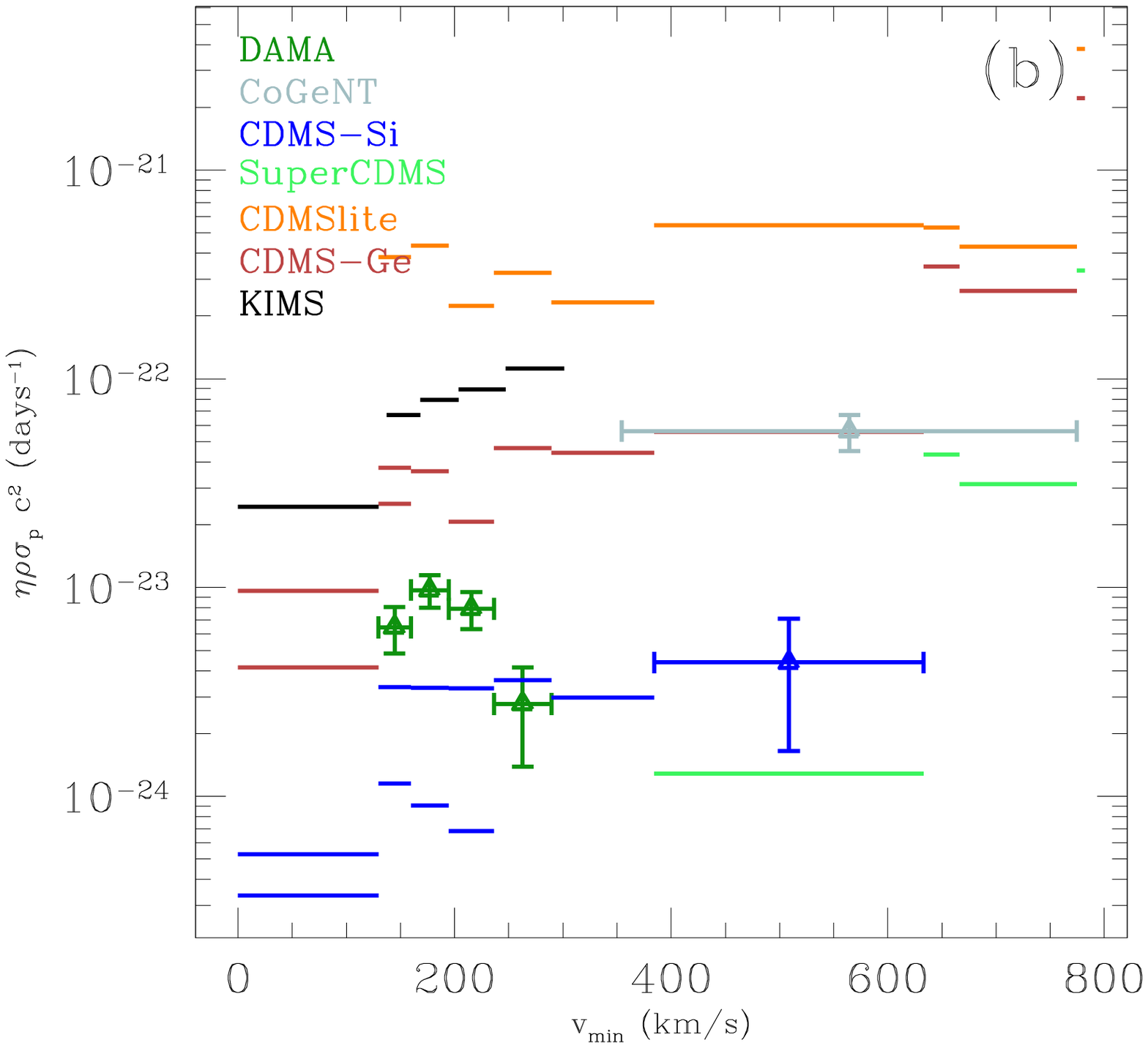}
\end{center}
\caption{Measurements and bounds for the functions $\tilde{\eta}_0$ and
  $\tilde{\eta}_1$ for the two benchmark points represented with
  crosses in Figs.\ref{fig:mchi_delta_i}(a-b) and for
  $f_n/f_p$=-0.69. {\bf (a)} $m_{DM}=$45 GeV, $\delta$=-37 keV; {\bf
    (b)} $m_{DM}=$300 GeV, $\delta$=-48 keV.}
\label{fig:benchmarks_na_quenching_collar}
\end{figure}

As discussed at the end of Section \ref{sec:shape_test}, for a
Maxwellian velocity distribution and when $f_n/f_p$=1, WIMPs heavier
than approximately 20 GeV interact in DAMA predominantly by scattering
off Iodine targets. However, if $f_p/f_n\ne$1 and when a wider class
of galactic velocity distributions is allowed, WIMP--$Na$ scatterings
can dominate also for heavier WIMP masses.  In this case DAMA and KIMS
can decouple also at large values of $m_{DM}$. Mainly motivated by
this possibility, in this Section we wish to explore this scenario in
more detail.

In Section \ref{sec:dama_na_low_mass} we already pointed out that
within all the $m_{DM}$--$\delta$ range of
Eq. (\ref{eq:mchi_delta_ranges}) we found
\verb!V_SUPERCDMS!$\le$\verb!V_DAMA!, so that the SuperCDMS
constraint could not be decoupled from the DAMA effect if scattering
off $Na$ nuclei was assumed. Nevertheless, that result was obtained by
assuming $Q_{Na}$=0.3 for the Sodium quenching factor, and it has to
be pointed out that the measurement of the latter is presently
somewhat controversial. In particular in Ref. \cite{quenching_collar}
a determination of $Q_{Na}$ significantly smaller than previous ones
is discussed. In this Section we will repeat the analysis of Section
\ref{sec:dama_na_low_mass} using this latter determination (see
Appendix \ref{app:exp} for details). The result is shown in
Fig. \ref{fig:mchi_delta_na_quenching_collar} where in the
horizontally hatched region \verb!V_DAMA_NA! $\subset$ \verb!V_GAL! ,
while, as a consequence of the different quenching factor, in the
+45$^{\circ}$ hatched domain now
\verb!V_SUPERCDMS!$>$\verb!V_DAMA_NA! is possible. However now we find
that \verb!V_LUX!$\le$\verb!V_DAMA_NA! over all the
(\ref{eq:mchi_delta_ranges}) ranges for the parameters. Nevertheless,
one can again assume $f_n/f_p\simeq$-0.69 to suppress the WIMP
coupling to Xenon. In this case this region of the parameter space can
be in agreement with both constraints. Notice that, as shown in
Fig. \ref{fig:scaling_laws}, this particular choice of $f_n/f_p$ has
also the effect of suppressing the couplings of WIMPs to Iodine and
Cesium compared to that to Sodium, so that in this range dominance of
scatterings off $Na$ can be consistently assumed in $NaI$ and also the
KIMS constraint can be relieved. Moreover, in
Fig. \ref{fig:mchi_delta_na_quenching_collar} the light (gray) shaded
region represents the parameter space where \verb!V_DAMA_NA!  $\cap$
\verb!V_DAMA_I!$\ne$0. As discussed in Section \ref{sec:shape_test},
no matter which WIMP--$Xe$ coupling suppression mechanism is assumed,
outside that band dominance of $Na$ (actually, of any of the two
targets) in DAMA can be assumed, albeit at the price of tuning the
$\tilde{\eta}_1(v_{min})$ to a sufficiently small value in the
$v_{min}$ ranges mapped by $I$. Notice that the same range of
$v_{min}$ would also interest scatterings of WIMPS off $CsI$ in part
of the energy range analyzed by KIMS, so that this would be another
way to (at least) reduce the tension between KIMS and DAMA.  Finally,
in Fig. \ref{fig:mchi_delta_na_quenching_collar} the (light--blue)
shaded area is excluded by the shape test.

The thick (red) boundary encloses the areas subject to all the
requirements and constraints.  Within those boundaries we select two
benchmark points, marked by crosses, that are analyzed in the
$v_{min}$=$\tilde{\eta}_{0,1}$ parameter space in
Figs. \ref{fig:benchmarks_na_quenching_collar}(a-b). Notice that in
Fig. \ref{fig:mchi_delta_i_w} it was possible to find a region in
the $m_{DM}$--$\delta$ parameter space where WIMPs dominantly scatter
off Tungsten nuclei in CRESST , in spite of the enhanced coupling to
Calcium for the choice $f_n/f_p\simeq$-0.69. This was possible because
in that Figure $\delta>$0: in that case for Calcium and Oxygen
$v^*_{min}$ is larger than for Tungsten, and driven beyond $v_{esc}$
for $\delta$ large enough, leaving only the contribution of $W$
nuclei. In the case of Fig. \ref{fig:mchi_delta_na_quenching_collar},
however, the region allowed by the SuperCDMS bound corresponds to
$\delta<0$. In this case $v_{min}^*$=0 and there is no longer a
clear--cut hierarchy among the $v_{min}$ ranges pertaining to the
three different targets. The same thing happens for the scaling--law
factors, which in CaWO$_4$ scale as $\tilde{A}_{Ca}$ : 4
$\times\tilde{A}_{O}$ : $\tilde{A}_{W}$ = 1 : 0.54 : 0.22.  This means
that a complicated pattern of dominances depending no only on $m_{DM}$
and $\delta$, but also on the recoil energy is expected in this case,
intertwined with domains of the parameter space where the
factorization introduced in Section \ref{sec:factorization} is not
possible in the first place. For this reason in
Figs. \ref{fig:benchmarks_na_quenching_collar}(a-b) CRESST is not
included in the discussion. Inspection of
Figs. \ref{fig:benchmarks_na_quenching_collar}(a-b) reveals that, as
expected, the discrepancy between DAMA and KIMS is relieved.
Moreover, as required, the SuperCDMS bound is no longer effective on
the DAMA points. Nevertheless a tensions develops in this case between
DAMA and the CDMS--$Si$ data
(Fig. \ref{fig:benchmarks_na_quenching_collar}(a)) or both the
CDMS--$Si$ and the CDMS--$Ge$ data
(Fig. \ref{fig:benchmarks_na_quenching_collar}(b)).

\section{Conclusions}

In the Inelastic Dark Matter scenario, the halo--model factorization
approach used to compare results from Dark Matter direct detection
experiments is more complicated that in the elastic case, because in
presence of a mass splitting $\delta\ne$0 the mapping between the
nuclear recoil energy $E_R$ and the minimal velocity $v_{min}$ that
the incoming WIMP needs to have to deposit $E_R$ becomes more involved
than in the elastic case.  For this reason a systematic analysis of
IDM where all available data are included making use of the
factorization property of the halo--model dependence was still missing
so far.  In the present paper we have attempted to address this issue,
introducing some strategies to determine regions in the IDM parameter
space where the tension existing among different experimental results
can be (at least partially) alleviated.

To this aim we have first introduced some internal consistency checks
involving the data of one single experiment, which exploit the fact
that, when the same $v_{min}$ range is mapped in two different energy
intervals, the expected correlation can be compared with the
data. Moreover, we have argued that, if a minimal set of assumptions
is adopted for the WIMP velocity distribution, the tension between the
putative signal from an experimental excess and the constraint from a
null result can be reduced or eliminated provided that the two results
can be mapped into non--overlapping ranges of $v_{min}$ and if the
$v_{min}$ range of the constraint is at higher values compared to that
of the excess. We stress that this latter argument involves
exclusively kinematics, and is valid no matter what the dynamics of
the process is.

We have then shown that, in the elastic case, the constraints from
XENON100, LUX and SuperCDMS are the most binding, and argued that this
hierarchy among limits is preserved in the IDM case. Then, adopting
the two criteria summarized above, we have systematically explored the
IDM parameter space to find regions where the XENON100, LUX and the
SuperCDMS constraints are relaxed, and picked within those regions
some representative benchmark points where we have discussed in more
detail the experimental situation including all the other bounds.

Following the strategy outlined above, we have then singled out five
scenarios:
\begin{itemize}

\item[i)] 2 GeV$\lsim m_{DM}\lsim$ 4 GeV, -130
  keV$\lsim\delta\lsim$-45 keV (see Fig. \ref{fig:mchi_delta_si}): in
  this approximate domain both an explanation of the DAMA modulation
  effect through WIMP--$Na$ scattering and the excess of three
  WIMP--candidate events observed by CDM-$Si$ can be brought in
  agreement with other bounds if some dynamical mechanism such as
  isospin violation can be advocated to suppress WIMP interaction with
  Germanium.  However, the DAMA and CDMS--$Si$ results turn out to be
  in mutual tension.

\item[ii)] $m_{DM}\gsim$ 60 GeV, 50 keV$\lsim\delta\lsim$ 180 keV; a
  wide band with $\delta\lsim$ -40 keV (see
  Fig.\ref{fig:mchi_delta_i}): in this approximate domain the tension
  between an explanation of the DAMA modulation effect in terms of
  WIMP--Iodine scattering can be alleviated by reducing it to that
  with the minimal number of other experiments: KIMS (which uses the
  same target nucleus) and XENON100 (which turns out to be more
  constraining than LUX thanks to the higher value of the upper bound
  of its analyzed energy region in keVnr). Notice that in the usual
  case when an Isothermal Sphere model for the velocity distribution
  is assumed, besides KIMS and XENON100 the DAMA region at large
  $m_{DM}$ appears to be well inside the domain excluded also by LUX
  and/or SuperCDMS (depending on the $\delta$ parameter). This may be
  interpreted as to strengthen the robustness of the exclusion, in
  spite of the many uncertainties existing in each experiment when
  taken separately. In our analysis we have shown that sometimes this
  argument can be misleading, and the number of experiments
  necessarily in tension with DAMA at large WIMP masses can be lower
  than generally assumed.

\item[iii)] $m_{DM}\gsim$ 30 GeV, 10 keV$\lsim\delta\lsim$ 100 keV
  (see Fig.\ref{fig:mchi_delta_w}) : in this approximate domain the
  excess measured by CRESST\cite{cresst} can be made compatible with
  all other constraints.

\item[iv)] $m_{DM}\gsim$ 350 GeV, 50 keV$\lsim\delta\lsim$ 130 keV
  (see Fig.\ref{fig:mchi_delta_i_w}). Mutual compatibility can be
  achieved between DAMA and CRESST in compliance with other
  constraints with the exception of KIMS. The size of the
  corresponding region in the $m_{DM}$--$\delta$ parameter space
  varies depending on the $f_n/f_p$ parameter.

\item[v)] If a measurement of the Sodium quenching factor
  substantially smaller compared to other measurements is
  adopted\cite{quenching_collar} and assuming a suppression
  mechanism for the WIMP--Xenon coupling, it is possible to single out
  a region of the parameter space with $m_{DM}\gsim$ 40 GeV, -50
  keV$\lsim\delta\lsim$ -20 keV where WIMP--Sodium scattering
  dominates in DAMA (see
  Fig. \ref{fig:mchi_delta_na_quenching_collar}). In this regime the
  SuperCDMS bound is evaded and also KIMS is not
  constraining. However, now DAMA appears in tension with CDMS-$Si$
  and CDMS-$Ge$.
\end{itemize}

All the compatibility regions listed above cannot be achieved if a
standard Isothermal Sphere is adopted for the WIMP velocity
distribution. 

We conclude by pointing out that direct detection experiments are
affected by many sources of possible systematic errors (including the
many uncertainties connected to quenching factors, atomic form
factors, background cuts efficiencies, etc.) that may affect
significantly the compatibility regions listed above. For instance, in
the specific example of Section \ref{sec:na_large_mchi} we have shown
that the adoption of a different measurement of the Sodium quenching
factor can lead to a very different scenario for the allowed parameter
space.

\acknowledgments 
This work was supported by the National Research
Foundation of Korea(NRF) grant funded by the Korea government(MOE)
(No. 2011-0024836).

\appendix
\section{Experimental inputs for the analysis}
\label{app:exp}

In this Appendix we summarize the experimental inputs that we have
used to evaluate the response function defined in
Eq.(\ref{eq:response_function}) for each of the experiments included
in our analysis. Whenever applicable we will follow the convention to
indicate with $E_R$ the true recoil energy, with $E_{ee}$ the
electron--equivalent energy ($E_{ee}=Q(E_R) E_R$ with $Q$ the
quenching factor) and with $E^{\prime}$ the visible energy, as
introduced in Section \ref{sec:factorization}. In the case of
bolometric measurements (CDMS, CRESST) we assume $Q=1$.  With the
exceptions of LUX and XENON100 we model the energy resolution with a
Gaussian and we indicate the corresponding variance.

{\bf DAMA} We have taken the modulation amplitudes in 0.5 keVee bins
from Fig.6 of Ref.\cite{dama} (already normalized to counts/day/kg/keV
for a total exposure of 1.17 ton yr), adopting the signal region 2
keVee $\le E^{\prime}\le$ 4 keVee in all plots and when discussing the
ranges of $v_{min}$ in Section
\ref{sec:phenomenology_inelastic}. Moreover, in order to perform the
mirror test introduced in Section \ref{sec:shape_test} we have used
the extended range 2 keVee$\le E^{\prime}\le$20 keVee.  We have
adopted the value $Q_I$=0.07 for the quenching factor for
Iodine\footnote{We choose $Q_I$=0.07 to maximize the compatibility
  regions shown in Figs. \ref{fig:mchi_delta_i} and
  \ref{fig:mchi_delta_w}. While presently the DAMA collaboration uses
  $Q_I$=0.09, the value $Q_I$=0.07 was adopted by DAMA in its early
  papers \protect \cite{bacci}, while a value as low as $Q_I$=0.05 is
  quoted in \protect \cite{fushimi}. Our choice is within the large
  systematic uncertainties on $Q_I$\cite{collar_quenching_i}.}, while
we have assumed two different determinations for the quenching factor
for Sodium: in Section \ref{sec:dama_na_low_mass} we have used
$Q_{Na}$=0.3, while in Section \ref{sec:na_large_mchi} we have adopted
the determination shown in Fig. 9 of Ref.\cite{quenching_collar}
fitting the experimental points with the functional form
$Q_{Na}(E_R)=0.024*\sqrt{E_R/\mbox{keVnr}}$. For the energy resolution
we have taken $\sigma_{DAMA}=0.0091 (E^{\prime}/\mbox{keVee})+0.448
\sqrt{E^{\prime}/\mbox{keVee}}$ in keVee.

{\bf XENON100 and LUX} In the case of LUX we have assumed zero WIMP
candidate events in the range 2 PE$\le S_1\le$30 PE in the lower half
of the signal band, as shown in Fig. 4 of Ref. \cite{lux} for the
primary scintillation signal $S_1$ (directly in Photo Electrons, PE)
for an exposure of 85.3 days and a fiducial volume of 118 kg of Xenon.
On the other hand for XENON100 we assumed the spectrum from Fig. 2 of
\cite{xenon100}, consisting in two events at $S_1$=[3.3 PE, 3.8 PE] in
the experimental range 3 PE$\le S_1\le$30 PE for an exposure of 224.6
days and a fiducial volume of 34 kg. In both cases, following
Ref. \cite{xenon100_response} (see Eqs. (14-15)) we have modeled the
detector's response with a Poissonian fluctuation of the $S_1$
scintillation photoelectrons combined with a Gaussian resolution
$\sigma_{PMT}$=0.5 PE for the photomultiplier so that the response
function defined in Eq.(\ref{eq:response_function}) is modified into:

\begin{eqnarray}
&&{\cal R}_{[S_{1,min},S_{1,max}]}=\frac{N_T m_N \tilde{A}^2}{2 \mu_{\chi{\cal N}}^2} F^2(E_R) MT \times\\
&&
\int_{S_{1,min}}^{S_{1,max}}dS_1 \, \sum_{n=1}^{\infty} Gauss\left (S_1 | n,\sqrt{n} \sigma_{PMT} \right ) 
Poiss\left [n,\nu(E_R) \right ]\xi_{cuts}(S_1).
\label{eq:response_function_liquid}
\end{eqnarray}
 
\noindent In the equation above
$Poiss(n,\lambda)=\lambda^n/n!\exp(-\lambda)$, while $\xi_{cuts}$
represents the combination of a 50\% acceptance combined with the
quality cut efficiency (taken from Fig. 9 of \cite{lux} for LUX and
from Fig.1 of \cite{xenon100} for XENON100). Moreover the expected
number of PE for a given recoil energy $E_R$ is given by:

\begin{equation}
  \nu(E_R)=E_{R}\times L_{eff}(E_R) \times L_y\frac{S_{nr}}{S_{ee}},
\label{eq:nu}
\end{equation}

\noindent with $L_y$=8.8 PE for LUX and $L_y$=2.28 PE for XENON100.
For LUX we have taken $L_{eff}(E_R)$ from \cite{lux_slides}
(where it is calculated including the effect of the electric field, so
that $S_{nr}=S_{ee}=1$, and assumed to vanish for $S_1<$ 3 PE), while for
XENON100 we have taken $L_{eff}(E_R)$ from \cite{xenon100_leff}
(in this case $S_{nr}$=0.95 and $S_{ee}$=0.58).

{\bf CoGeNT} We do not consider the annual modulation but only the
spectral excess and take both the total count rates and the background
from Fig. 23 of \cite{cogent_spectral}\footnote{We adopt the
  subtraction of background surface events from the official data
  analysis of the CoGeNT Collaboration\cite{cogent_spectral}.  For a
  critical independent assessment of the CoGeNT spectral excess,
  claiming a much less significant residual effect than the official
  analysis, see \cite{cogent_davis}.} rescaling them to the latest
exposure of 1129 days of Ref. \cite{cogent_last} for a fiducial mass
of 0.33 kg of Germanium and assuming the signal range 0.5
keVee$<E_{ee}<$2 keVee. For the quenching factor we assume $Q_{Ge}=0.2
(E/\mbox{keVee})^{0.12}$ as given in \cite{cogent_spectral} while for
the energy resolution we use
$\sigma_{CoGeNT}(E^{\prime})=\sqrt{69.7^2+0.976 (E^{\prime}/eV)}$ in
eV \cite{cogent_resolution}.

{\bf CDMS-$Si$} We take the full energy range 7 keVnr$<E_R<$100 keVnr
analyzed in \cite{cdms_si} with an exposure of 140.2 kg day with a
Silicon target. The three WIMP candidate events are observed at
energies $E_R$=8.2 keVnr, 9.5 keVnr and 12.3 keVnr, so when discussing
the ranges $v_{min}$ in Section \ref{sec:dama_na_low_mass} or the
consistency checks in Section \ref{sec:shape_test} we define the
signal region 8 keVnr$<E_R<$12.5 keVnr. Since the energy resolution in
CDMS-$Si$ has not been measured we take
$\sigma_{CDMS-Si}(E^{\prime})=\sqrt{0.293^2+0.056^2
  (E^{\prime}/\mbox{keVnr})}$ in keVnr from \cite{cdms_resolution}.

{\bf SuperCDMS} We include the low--energy analysis of
SuperCDMS\cite{super_cdms} with a Germanium target in the energy
range 1.6 keVnr$<E_R<$ 10 keVnr with a total exposition of 577 kg day
and 11 observed WIMP candidates. The energy resolution is given by
$\sigma_{CDMS-Si}(E^{\prime})=\sqrt{0.293^2+0.056^2
  (E^{\prime}/\mbox{keVnr})}$ in keVnr\cite{cdms_resolution}

{\bf XENON10} The analysis of XENON10 makes use of the secondary
ionization signal $S_2$ only, with an exposition of 12.5 day and a
fiducial mass of 1.2 kg. We take the scale of the recoil energy $E_R$
and the recorded event spectrum in the energy range 1.4 keVnr$<E_R<$
10 keVnr directly from Fig. 2 of Ref. \cite{xenon10}. The energy
resolution is given by: $\sigma_{XENON10}=E_R/\sqrt{E_R Q_y(E_R)}$
where $Q_y(E_R)$ is the electron yield that we calculate with the same
choice of parameters as in Fig. 1 of \cite{xenon10}.

{\bf CDMSlite} CDMSlite\cite{cdms_lite} analyzes the very low range
0.170 keVee$<E_{ee}<$7 keVee for the electron--equivalent energy using a
fiducial mass of 0.6 kg of Germanium and an exposition of 10.3
days. We take the spectrum from Fig. 1 of Ref. \cite{cdms_lite}. We
adopt the same quenching factor that we use for CoGeNT, an energy
resolution $\sigma_{CDMSlite}=$14 eV and the efficiency
$\xi_{cut}=$0.985 \cite{cdms_lite}.

{\bf CDMS-$Ge$} We consider the data from detector T1Z5 in the range 2
keVnr$<E_R<$ 100 keVnr available in digital format from \cite{cdms_ge}
with a raw exposure of 35 kg day on Germanium target. The energy
resolution is the same as in SuperCDMS, while the efficiency is taken
from Fig.1 of Ref. \cite{cdms_ge}.

{\bf CRESST} We only focus on scatterings on Tungsten in CaWO$_4$. To
this aim we select from \cite{cresst} the 45 events (out of 67) in the
$W$ recoil bands of Figs. 7, 9 and 17 in the total energy range 10
keVnr$<E_R<$ 40 keVnr collected with an exposition of 730 kg day
\footnote{After submission of the present manuscript new unpublished
  CRESST data have been presented in \protect\cite{tevpa_idm_cresst}
  that do not confirm the excess claimed in \cite{cresst}.}. The
background in the $W$ band is dominated by lead recoils from
$^{210}$Po decays, which we model as in Eq. (1) of
Ref. \cite{cresst}. When discussing $v_{min}$ ranges in Section
\ref{sec:phenomenology_inelastic} and the consistency checks of
Section \ref{sec:shape_test} we select the signal region 12 keVnr
$<E_R<$ 24 keVnr where we optimize the signal/background ratio getting
34 total events vs. a background of 7.4. As far as the energy
resolution is concerned, we use the two measurements
FWHM(E$^{\prime}$=3.6 keVnr) = 0.3 keV and FWHM(E$^{\prime}$= 64
keVnr) = 1.6 keVnr from \cite{cresst_slides} to fit the functional
form $\sigma_{CRESST}(E^{\prime})=$-0.0442+0.0904$\sqrt{E^{\prime}}$
in keVnr (FWHD=2.355$\times \sigma$).

{\bf KIMS} We take the 90\% C.L. upper bounds on nuclear recoil events
from Fig. 4 of Ref. \cite{kims} (for 3 keVee$<E^{\prime}<$11 keVee in
1--keVee bins, already in counts/day/kg/keV for an effective exposition
of 24524.3 kg day) rebinning them using Eq.(\ref{eq:dama_rebin}). We
use as quenching factor the solid line in Fig.13 of
Ref. \cite{kims_quenching_factor} (in the measurement the quenching
factors of $Cs$ and $I$ cannot be distinguished and are assumed to be
the same). For the energy resolution we have rescaled the
FWHM$\simeq$14.24 keVee of the peak at 59.5 keVee from $^{241}Am$
calibration shown if Fig. 8 of Ref.\cite{kims_resolution}, getting
$\sigma_{KIMS}(E^{\prime})=0.78 \sqrt{E^{\prime}}$ in keVee. Since the
atomic numbers of Iodine ($A$=127) and Cesium ($A=133$) are very close,
both contributions can be incorporated in the definition of the
response function, i.e.

\begin{equation}
  \bar{\tilde{\eta}}=\frac{\int_{0}^{\infty} d v_{min} \tilde{\eta}(v_{min})\left [ {\cal R}^{Iodine}_{[E_1^{\prime},E_2^{\prime}]}(v_{min})+{\cal R}^{Cesium}_{[E_1^{\prime},E_2^{\prime}]}(v_{min})\right ]}
  {\int_{0}^{\infty} d v_{min} \left [{\cal R}^{Iodine}_{[E_1^{\prime},E_2^{\prime}]}(v_{min})+{\cal R}^{Cesium}_{[E_1^{\prime},E_2^{\prime}]}(v_{min})\right]}.
\label{eq:eta_bar_vmin_csi}
\end{equation}

\noindent This of course can be done for any multi--target detector,
but makes sense only when two response functions largely overlap, as
in the case of $CsI$. The net effect is to somehow worsen in $v_{min}$
space the smearing effect of the energy resolution, but is practically
irrelevant (we checked that it amounts to less than 3\% on $v_{min}$
in all the benchmarks plotted). In particular, in the case $f_n/f_p=1$
one has $\tilde{A}_{Cesium}/\tilde{A}_{Iodine}\simeq 1.1$, and the
ensuing $v_{min}$ range is given by the combination of the (slightly
offset) Cesium and Iodine mappings from the recoil energy to
$v_{min}$. However, in the case $f_n/f_p=-0.69$ discussed in
Figs. \ref{fig:benchmark_mchi_45_delta_70} and
\ref{fig:benchmarks_na_quenching_collar}
$\tilde{A}_{Cesium}/\tilde{A}_{Iodine}\simeq 0.15$ (see
Fig. \ref{fig:scaling_laws}): in that particular case Iodine can be
taken as the dominant target and the resulting $v_{min}$ ranges
correspond to WIMP--Iodine scattering only.

\end{document}